\documentclass{aa}
\usepackage{graphicx}
\usepackage{natbib}
\usepackage{bm,times,color,url}
\graphicspath{{./fig/}{./png/}}

\newcommand{\EQ}{\begin{equation}}
\newcommand{\EN}{\end{equation}}
\newcommand{\EQA}{\begin{eqnarray}}
\newcommand{\ENA}{\end{eqnarray}}

\newcommand{\Eq}[1]{Eq.~(\ref{#1})}

\newcommand{\Sec}[1]{Sect.~\ref{#1}}

\newcommand{\Fig}[1]{Figure~\ref{#1}}
\newcommand{\FFig}[1]{Figure~\ref{#1}}

\newcommand{\Figs}[2]{Figures~\ref{#1} and \ref{#2}}

\newcommand{\Tab}[1]{Table~\ref{#1}}

\newcommand{\bra}[1]{\langle #1\rangle}

\newcommand{\meanrho}{\overline{\rho}}

{}
{}
{}

\newcommand{\meanSSSS}{\overline{\mbox{\boldmath ${\mathsf S}$}} {}}

\newcommand{\meanSSS}{\overline{\mathsf{S}}}
{}
{}
{}
{}
{}
{}
{}
{}
\newcommand{\meanEE}{\overline{\mbox{\boldmath $E$}}{}}{}
{}
{}
{}
{}
{}
{}
{}
{}
\newcommand{\meanUU}{\overline{\bm{U}}}

{}

\newcommand{\meanA}{\overline{A}}
\newcommand{\meanB}{\overline{B}}

\newcommand{\meanU}{\overline{U}}

\newcommand{\meanp}{\overline{p}}

{}

{}
{}
{}

\newcommand{\pphi}{\hat{\bm{\phi}}}

\newcommand{\rrr}{\hat{\mbox{\boldmath $r$}} {}}

\newcommand{\meanAA}{{\overline{\bm{A}}}}
\newcommand{\meanBB}{{\overline{\bm{B}}}}
\newcommand{\meanJJ}{{\overline{\bm{J}}}}

\newcommand{\rr}{\bm{r}}

\newcommand{\BB}{\bm{B}}

\newcommand{\FF}{\bm{F}}

\newcommand{\SSS}{\bm{S}}

\newcommand{\nab}{{\bm{\nabla}}}

\newcommand{\meanQQQ}{\overline{\mbox{\boldmath ${\cal Q}$}} {}}
\newcommand{\DD}{{\rm D} {}}

\newcommand{\const}{{\rm const}  {}}

\def\degr{\hbox{$^\circ$}}

\def\cp{c_{\rm p}}

\def\cs{c_{\rm s}}
\def\rcrit{r_{\rm c}}

\def\vA{v_{\rm A}}

\def\urms{u_{\rm rms}}

\def\nuT{\nu_{\rm T}}

\def\etaT{\eta_{\rm T}}

\def\Beq{B_{\rm eq}}

\def\half{{\textstyle{1\over2}}}

\def\onethird{{\textstyle{1\over3}}}

\newcommand{\W}{\,{\rm W}}

\newcommand{\G}{\,{\rm G}}

\newcommand{\K}{\,{\rm K}}
\newcommand{\g}{\,{\rm g}}
\newcommand{\s}{\,{\rm s}}

\newcommand{\cm}{\,{\rm cm}}

\newcommand{\km}{\,{\rm km}}

\newcommand{\kms}{\,{\rm km\,s}^{-1}}

\newcommand{\yr}{\,{\rm yr}}

\newcommand{\AU}{\,{\rm AU}}

\newcommand{\yjpp}[3]{ #1, {J. Plasma Phys.,} {#2}, #3}
\newcommand{\yapj}[3]{ #1, {ApJ,} {#2}, #3}

\newcommand{\yapjl}[3]{ #1, {ApJ,} {#2}, #3}

\newcommand{\yzfa}[3]{ #1, {Z.\ f.\ Ap.,} {#2}, #3}

\newcommand{\yana}[3]{ #1, {A\&A,} {#2}, #3}

\newcommand{\ygafd}[3]{ #1, {Geophys.\ Astrophys.\ Fluid Dyn.,} {#2}, #3}

\newcommand{\yjfm}[3]{ #1, {J.\ Fluid Mech.,} {#2}, #3}

\newcommand{\yptrs}[3]{ #1, {Phil.\ Trans.\ R.\ Soc.,} {#2}, #3}

\newcommand{\ymn}[3]{ #1, {MNRAS,} {#2}, #3}
\newcommand{\ynat}[3]{ #1, {Nature,} {#2}, #3}

\newcommand{\ysph}[3]{ #1, {Solar Phys.,} {#2}, #3}

\newcommand{\yjour}[4]{ #1, {#2}, {#3}, #4}

\newcommand{\ybook}[3]{ #1, {#2} (#3)}
\newcommand{\yproc}[5]{ #1, in {#3}, ed.\ #4 (#5), #2}

\newcommand{\pana}[2]{ #1, {A\&A}, in press, arXiv:#2}

\def\apjs{ApJS}

\titlerunning{}
\authorrunning{P. Jakab and A. Brandenburg}
\title{The effect of a dynamo-generated field on the Parker wind}
\author{P. Jakab\inst{1,2} and A. Brandenburg\inst{1,3,4,5}
}
\institute{
Nordita, KTH Royal Institute of Technology and Stockholm University,
Roslagstullsbacken 23, 10691 Stockholm, Sweden
\and
Department of Nuclear and Subnuclear Physics,
Pavol Jozef Safarik University in Kosice, 
Srobarova 2, 041 54 Kosice, Slovakia 
\and
Department of Astronomy, AlbaNova University Center,
Stockholm University, 10691 Stockholm, Sweden
\and
JILA and Laboratory for Atmospheric and Space Physics,
University of Colorado, Boulder, CO 80303, USA
\and
McWilliams Center for Cosmology \& Department of Physics,
Carnegie Mellon University, Pittsburgh, PA 15213, USA
}
\date{\today,~ $ $Revision: 1.101 $ $}
\begin{document}

\abstract{
Stellar winds are an integral part of the underlying dynamo, the motor
of stellar activity.
The wind controls the star's angular momentum loss, which depends on the
magnetic field geometry which, in turn, varies significantly in time
and latitude.
}{
Here we study basic properties of a self-consistent model that includes
simple representations of both the global stellar dynamo in a spherical
shell and the exterior in which the wind accelerates and becomes
supersonic.
}{
We numerically solved an axisymmetric mean-field model for the induction,
momentum, and continuity equations using an isothermal equation of state.
The model allows for the simultaneous generation of a mean magnetic
field and the development of a Parker wind.
The resulting flow is transonic at the critical point, which we arranged
to be between the inner and outer radii of the model.
The boundary conditions are assumed to be such that the magnetic field
is antisymmetric about the equator, that is to say dipolar.
}{
At the solar rotation rate, the dynamo is oscillatory and of
$\alpha^2$ type.
In most of the domain, the magnetic field corresponds to that of
a split monopole.
The magnetic energy flux is largest between the stellar surface and
the critical point.
The angular momentum flux is highly variable in time and can reach
negative values, especially at midlatitudes.
At a rapid rotation of up to 50 times the solar value, most of the
magnetic field is lost along the axis within the inner tangential
cylinder of the model.
}{
The model reveals unexpected features that are not generally anticipated
from models that are designed to reproduce the solar wind: highly
variable angular momentum fluxes even from just an $\alpha^2$
dynamo in the star.
A major caveat of our isothermal models with a magnetic field produced
by a dynamo is the difficulty to reach small enough plasma betas without
the dynamo itself becoming unrealistically strong inside the star.
}

\maketitle

\section{Introduction}
The emergence of a wind around stars is a remarkable and somewhat
counter-intuitive phenomenon.
The existence of the solar wind was already suggested because
the tails of comets always point away from the Sun \citep{Bie51}.
Nevertheless, the wind was thought to be a relatively slow phenomenon
associated with evaporation of the corona \citep{Cha60}.
The physical nature and mathematical theory of the solar wind was first
understood by \cite{Par58}.
His theory showed that the wind starts off as a subsonic flow some
distance above the corona.
It gradually gains in speed as the gravitational force diminishes and
the effective outward pull resulting from the quadratic increase of the
cross-sectional area in Bernoulli's law becomes dominant.
This is a purely hydrodynamic phenomenon, unlike what was suggested by
the popular notion of the solar corpuscular radiation at the time.

Stellar winds play a crucial role in a star's life.
Without the wind, the Sun would still be spinning rapidly and
magnetically superactive.
A proper understanding of the rotational evolution of a star through
magnetic braking via a wind is important not only for stellar evolution,
but it also plays a role in understanding the diversity of magnetic
activity as a function of the rotation rate and age \citep{vanSaders2016}.
As the star reaches the age of the Sun, the magnetic field either changes
its geometry such that stellar braking is reduced \citep{MvS17,See19}
or it can continue to brake and the star's differential rotation becomes
antisolar-like \citep{GYMRW14,KKB14}, that is, the equator spins slower
than the poles.
Stellar winds can also be important for the dynamo itself in that they
can transport magnetic helicity away from the dynamo region, and thereby
alleviate what is known as catastrophic quenching; see \cite{MMTB11}
for mean-field models and \cite{DSGB13} for computations of the magnetic
helicity flux in simulations in a turbulent wind.
Magnetic winds also affect the density and dynamics of cosmic rays in
the heliosphere.
Selfconsistently computing the dynamo-generated magnetic field evolution
in the heliosphere is, therefore, also crucial for modeling the magnetic
shielding of Galactic cosmic rays on Earth.

The theory of a magnetized stellar wind by \cite{WD67} employed a
prescribed and time-independent stellar magnetic field, so any
feedback on the underlying dynamics was ignored.
This is also true of the recent numerical models of \cite{Rev15},
who compared different magnetic multipoles as initial conditions
of their models.
This has changed only in recent years.
Given that the wind normally dominates over the magnetic field, one can
separate the dynamics of the wind from that of the solar dynamo.
\cite{Pinto}, and more recently \cite{PBRS18}, modeled this by
using two separate codes that are magnetically coupled through
a matching condition at the solar surface.
In more recent work, \cite{PBRS20} extended their model to also include
a mean-field dynamo solution in the Pluto code, rather than matching
the solutions of two separate codes.
This allows for feedback from the wind onto the dynamo.
This is therefore similar to the work presented here, except that
they still invoke what they call a multilayered boundary condition.
This means that different equations are being solved inside and outside
the star.
The model is therefore still not fully self-consistent, but in some ways
more realistic than ours.

The purpose of the present paper is to explore some basic properties of
stellar winds in the presence of dynamo-generated magnetic fields.
It is appropriate to adopt a mean-field model, where we solve the
equations for the azimuthally averaged magnetic and velocity fields.
In this paper, those mean fields are denoted by an overbar.
The effects of turbulence are then parameterized through a
turbulent viscosity and a turbulent magnetic diffusivity.
In the star's convection zone, there are also cyclonic convective motions
giving rise to kinetic helicity of opposite signs in the two hemispheres.
This is modeled through an $\alpha$ effect \citep{KR80}.
The turbulent magnetic diffusivity is here assumed constant.

The presence of the magnetic field causes the kinetic and magnetic
stresses to be different from zero.
The turbulent viscosity is itself a result of kinetic and magnetic
stresses caused by the fluctuating components of the magnetic and
velocity fields.
In the theory of turbulent accretion disks \citep{FKR92}, those stresses
are parameterized by the \cite{SS73} parameter, $\alpha_{\rm SS}$.
It quantifies the stress in terms of the background differential rotation,
the sound speed, and the scale height.
In accretion disks, where the differential rotation is Keplerian,
this amounts to a scaling of the stress by the sound speed squared.
In our case, the differential rotation is not related to the sound speed,
but the basic mechanism of angular momentum transfer is the same, and
we can still express the total stress in a similar fashion.

Unlike the work of \cite{PBRS18}, we consider the evolution of the dynamo
and the wind within a single code.
At this point, our aim is not to produce a realistic model of the Sun,
but rather a physically consistent model under conditions where the
dynamics of the wind can no longer be separated from that of the dynamo.
Our models can also be applied to conditions of rapid rotation, which
strongly affects the wind.
This can be particularly relevant to young stars in their T Tauri phase.
We begin by presenting the basic equations of our model and turn then to
the discussion of our results.

The simplest wind solution is the isothermal one that was already found
by \cite{Par58}.
Heating is not explicitly invoked.
Its physics resembles that of a siphon flow.
Once a fluid parcel has moved over the top of the effective gravitational
potential, it simply continues to fall and pulls the remaining fluid
behind it \citep{Shore}.
The top of the effective potential corresponds to the critical point
where the flow speed crosses the sonic point.
We arrange this point to be in the middle of the computational domain
such that the flow speed becomes supersonic well before the outer point
$r_{\rm out}$.
We fit the dynamo-active zone (or stellar envelope) with an $\alpha$
effect different from zero into a spherical shell between the inner
point of the computational domain, $r_{\rm in}$, and a radius $R$,
which models the surface of the star.

The usefulness of an isothermal solution can be justified by
considering the fact that the sound speed both at the bottom of the convection
zone and in the solar wind is about $100\kms$, corresponding to a
temperature of a million degrees.
The lower temperature near the photosphere is obviously ignored.
For an isothermal gas, the mean pressure $\meanp$ is then simply
proportional to the gas density $\meanrho$ with $\meanp=\meanrho\cs^2$,
where $\cs$ is the isothermal sound speed.
The pressure gradient is then given by
$(\nab\meanp)/\meanrho=\cs^2\nab\ln\meanrho$.
The implications of a cool photosphere will be discussed at the
end of the paper.

We begin by discussing first the basic equations, boundary conditions,
and parameters in \Sec{Model}.
We then present our results in \Sec{Results}, and draw our conclusions
in \Sec{Conclusions}.

\section{The model}
\label{Model}

We adopt spherical polar coordinates, $(r,\theta,\phi)$, with the origin
at the center of the star.
The vector $\rr$ points away from the center, the colatitude $\theta$
increases away from the north pole, and $\phi$ increases in the eastward
direction.
We assume axisymmetry, that is, $\partial/\partial\phi=0$.

\subsection{Basic equations}

We write the mean magnetic field as $\meanBB=\nab\times\meanAA$,
where $\meanAA$ is the mean vector potential.
This ensures that $\nab\cdot\meanBB=0$ at all times.
The evolution equations for $\meanAA$, the mean velocity $\meanUU$,
and the logarithmic mean density $\meanrho$, are
\EQA
\label{dAmean}
{\partial\meanAA\over\partial t}&=&\meanUU\times\meanBB
+\alpha\meanBB-\etaT\mu_0\meanJJ,\\
\label{dUmean}
{\DD\meanUU\over\DD t}&=&-\cs^2\nab\ln\meanrho
-{GM\over r^2}\rrr+{1\over\meanrho}\meanJJ\times\meanBB
-\nuT\meanQQQ,\\
{\DD\ln\meanrho\over\DD t}&=&-\nab\cdot\meanUU,
\ENA
where $\DD/\DD t=\partial/\partial t+\meanUU\cdot\nab$
is the advective derivative,
$G$ is Newton's constant, $M$ is the stellar mass,
$\rrr=\rr/r$ is the radial unit vector,
$\etaT$ and $\nuT$ are the sums of turbulent and
microphysical values of magnetic diffusivity and kinematic viscosity,
respectively, $\alpha$ is the aforementioned coefficient
in the $\alpha$ effect,
$\meanJJ=\nab\times\meanBB/\mu_0$ is the mean current density,
$\mu_0$ is the vacuum permeability,
\EQ
-\meanQQQ=\nabla^2\meanUU+\onethird\nab\nab\cdot\meanUU
+2\meanSSSS\cdot\nab\ln(\nuT\meanrho)
\label{calQ}
\EN
is a term appearing in the viscous force, where
$\meanSSSS$ is the traceless rate of strain tensor of the mean flow
with components $\meanSSS_{ij}=\half(\meanU_{i,j}+\meanU_{j,i})
-\onethird\delta_{ij}\nab\cdot\meanUU$.
The dot in \Eq{calQ} denotes the contraction over the free index
of $\nab\ln(\nuT\meanrho)$.

The mean magnetic field is generated by the $\alpha$ effect.
This leads to exponential growth, provided the value of $\alpha$ is
above a certain critical value.
Eventually, the dynamo must saturate 
because the Lorentz force from the mean field, $\meanJJ\times\meanBB$,
drives fluid motions that feed back onto the dynamo to limit its growth.
This way of achieving saturation is sometimes referred to as \cite{MP75}
mechanism.
In addition, there can be feedback from the small-scale magnetic field
that leads to a nonlinear suppression of $\alpha$, which is referred to
as $\alpha$ quenching.
We assume here a simple quenching function for $\alpha$, which is then
written in the form
\begin{equation}
\alpha(r,\theta,\meanBB)=\frac{\alpha_0 f_\alpha(r)\cos\theta\sin^n\theta}
{1+Q_\alpha\meanBB^2/\Beq^2},
\label{AlphaEqn}
\end{equation}
where $n=6$ is chosen to concentrate the $\alpha$ effect to low
latitudes \citep{JBKMR15,CBKK16}, $Q_\alpha$ is a quenching parameter
that determines the typical field strength, which is expected to be on
the order of $Q_\alpha^{-1/2}\Beq$, and
\begin{equation}
f_\alpha(r)=\Theta\Big((r-R)/w_\alpha\Big)
\label{AlphaProf}
\end{equation}
is a radial profile function with $\Theta(x)$ being
a smoothened step function from 0 to 1 as $x$ crosses zero.
Here, $R$ and $w_\alpha$ determine the location and
width of the transition.
The value of $Q_\alpha$ determines the nonlinear equilibration
of the dynamo, in addition to the macroscopic feedback from
the Lorentz force mentioned above.
Our model thus comprises three distinct layers with
\EQ
r_{\rm in}<R<\rcrit<r_{\rm out},
\EN
where $r_{\rm in}<r<R$ is the dynamo region (modeling the stellar envelope),
$R<r<\rcrit$ is the wind acceleration region (modeling the locations of the
solar corona and the Alfv{\'e}n point), and $\rcrit<r<r_{\rm out}$ is the
supersonic wind region with $\rcrit=GM/2\cs^2$ being the critical point.

\subsection{Boundary conditions}

In most of the cases, we apply a uniform angular velocity
$\Omega_0$ on the inner boundary $r=r_{\rm in}$ by setting
$u_\phi=r_{\rm in}\sin\theta\,\Omega_0$.
For the other two velocity components, we adopt ``open'' boundary
conditions by setting the second radial derivative to zero.
This condition turns out to be stable in all cases considered
in this paper.
It allows for a weak inflow to replenish the mass loss on
the outer boundary $r=r_{\rm out}$, where we apply open boundary
conditions for all three velocity components.
No precautions are taken to ensure that the mass in the computational
domain stays constant.
It turns out, however, that the total mass remains nearly unchanged.
This is, to some extent, also explained by the fact that the total
mass loss rate is small compared with other inverse time scales
in the problem.

For the magnetic field, we adopt a perfect conductor boundary condition
on the inner radius, that is,
\EQ
{\partial\meanA_r\over\partial r}=\meanA_\theta=\meanA_\phi=0
\quad\mbox{on $r=r_{\rm in}$},
\EN
and a radial field condition on the outer radius, that is,
\EQ
\meanA_r={\partial\meanA_\theta\over\partial r}+{\meanA_\theta\over r}
={\partial\meanA_\phi\over\partial r}+{\meanA_\phi\over r}=0
\quad\mbox{on $r=r_{\rm out}$}.
\EN
On the pole, we assume
\EQ
{\partial\meanA_r\over\partial\theta}
=\meanA_\theta=\meanA_\phi=0
\quad\mbox{on $\;\theta=0\degr\;$},
\EN
while on the equator, we assume
\EQ
{\partial\meanA_r\over\partial\theta}
=\meanA_\theta={\partial\meanA_\phi\over\partial\theta}=0
\quad\mbox{on $\;\theta=90\degr$}.
\EN
Since our simulations are axisymmetric, the magnetic field is
conveniently represented via $\meanB_\phi$ and $\meanA_\phi$.
In particular, contours of $r\sin\theta\,\meanA_\phi$ give the
magnetic field lines of the poloidal field,
$\meanBB_{\rm pol}=\nab\times(\meanA_\phi\pphi)$.

\subsection{Wind solution as initial condition}

As initial condition for $\meanUU\equiv(u,0,0)$ and $\meanrho$,
we adopt the Parker wind solution.
In some cases we also add a finite angular velocity with
constant angular momentum, although its effect on the
dynamics is ignored in the initial condition.
We begin by discussing the Parker wind solution,
which can be obtained by solving the Bernoulli equation,
\EQ
\half u^2 + \cs^2\ln\meanrho - GM/r=\const,
\EN
along with the equation of mass conservation,
which states that the mass loss rate is given by
$\dot{M}=4\pi r^2\meanrho u$.
We then obtain
\EQ
\half u^2 - \cs^2\ln u -\cs^2\ln r^2- GM/r=\Phi_0,
\EN
where $\Phi_0=-3/2$ is obtained by inserting the values
$u=\rcrit=1$ for the critical point.
We solve the Bernoulli equation iteratively.
For $r\leq \rcrit$, using $u=\cs r/\rcrit$ initially, we iterate
\EQ
\cs^2\ln u_{i+1}(r)=\half u_i^2-\cs^2\ln r^2-GM/r-\Phi_0,
\EN
while for $r>\rcrit$, using $u_0=2\cs$ initially, we iterate
\EQ
\half u^2_{i+1}(r)=\cs^2\ln u_i+\cs^2\ln r^2+GM/r+\Phi_0.
\EN
This iteration procedure was implemented by J\"orn Warnecke
and Dhrubaditya Mitra into the {\sc Pencil Code}\footnote{
\url{http://github.com/pencil-code} \citep{PCcollab},
DOI:10.5281/zenodo.2315093} in 2012.
We choose the initial value of $\dot{M}$ to be $\dot{M}_0$.

\subsection{Parameters and estimates for the Sun}

It is convenient to work with nondimensional units
by measuring speeds in units of the isothermal sound speed
and lengths in units of the critical radius, $\rcrit=GM/2\cs^2$.
In the following, we use tildae to denote nondimensional quantities.
Using typical numbers for the Sun, we have
\EQ
\cs=10^7\cm\s^{-1}=100\km\s^{-1},
\EN
$GM=GM_\odot\approx1.3\times10^{26}\cm^3\s^{-2}$, and therefore
\EQ
\rcrit=GM_\odot/2\cs^2\approx7\times10^{11}\cm
\approx10 R_\odot\approx0.05\AU.
\EN
In the Sun, the turbulent viscosity is
$\nuT\approx\urms\ell/3\approx10^{13}\cm^2\s^{-1}$.
The nondimensional viscosity is then
\EQ
\tilde{\nu}_{\rm T}\equiv{2\nuT\cs\over GM_\odot}\approx2\times10^{-6},
\EN
which is rather small.

For numerical stability, as already alluded to, we cannot choose the
value of $\nuT$ to be too small.
In practice, for a numerical resolution of $128\times32$ mesh points
in the $r$ and $\theta$ directions,
we can choose $\tilde{\nu}_{\rm T}\approx0.01$.
For $4096\times1024$ mesh points, on the other hand, we can reduce it
by a factor of 128 to $\tilde{\nu}_{\rm T}\approx8\times10^{-5}$.
This then also means that in the stellar convection zone, we cannot
adopt significantly smaller values, as is expected theoretically
based on our earlier estimates of $\urms$ and $\ell$.

The nondimensional value of the angular velocity is given by
\EQ
\tilde{\Omega}=\rcrit\Omega_0/\cs=GM\Omega_0/2\cs^3\approx0.2,
\EN
where we have used $\Omega_0=3\times10^{-6}\s^{-1}$.
The strength of the dynamo is determined by the two dynamo numbers,
\EQ
C_\alpha=\alpha_0 R/\etaT\quad\mbox{and}\quad
C_\Omega=\Delta\Omega R^2/\etaT.
\label{CalpCom}
\EN
where $\Delta\Omega$ is the angular velocity difference in the
equatorial plane of the stellar envelope.
The excitation conditions for dipolar and quadrupolar parities
are generally fairly close together \citep{Rob72}.
This is because the magnetic field is strongest at high latitudes,
so the hemispheric coupling is weak.
In the following we restrict ourselves to solutions with dipolar parity.
We vary the value of $C_\alpha$ and focus on values that are about
twice supercritical.

In our simulations, we adopt nondimensional units by setting
\EQ
\rcrit=\cs=\dot{M}_0=\mu_0=1,
\EN
which implies that $GM=2$.
Our unit of mass is then $[M]=\dot{M}_0 \rcrit/\cs$.
For the Sun, we have $\dot{M}_0\approx6\times10^{12}\g\s^{-1}$,
so that our unit of density is $[\rho]=\dot{M}_0/\cs \rcrit^2$,
which is about $1.2\times10^{-18}\g\cm^{-3}$ for the Sun.
Therefore, our unit of $\BB$ is $[B]=(\mu_0[\rho])^{1/2}\cs$,
which is about $0.04\G$ for the Sun.
The value of Newton's constant $G$ never enters on its own.
It could be determined a posteriori, if we knew the total stellar mass.
In our model, we can compute the mass $M_\ast$ of the stellar envelope
in $r_{\rm in}\leq r\leq R$, but this still leaves the mass of the
stellar core undetermined.
In the following, it is often convenient to retain the symbols $\rcrit$,
$\cs$, $\dot{M}_0$, and $\mu_0$ to remind ourselves of the normalization.

There are a few other parameters of the model that we keep fixed.
In all cases we use $w_\alpha=0.02$ for the transition thickness
of $\alpha$ near the surface; see \Eq{AlphaProf}.
We always take $r_{\rm in}=0.1$ and $R=0.2$.
This corresponds to a fractional shell thickness of 50\% instead of
the 30\% in the case of the Sun, but we should keep in mind that there
are other properties that agree with the Sun only qualitatively.
Another example is our smaller choice of $R/r_{\rm in}=5$ instead
of the solar value of about 10.
In all our simulations with $4096\times1024$ meshpoints,
we use $\tilde\nu_{\rm T}=8\times10^{-5}$.

\subsection{Comparison of characteristic time scales}
\label{TimeScales}

In our simulations, sound speed and the critical radius are set to unity,
so the characteristic sound travel time,
\EQ
\tau_{\rm s}=\rcrit/\cs
\EN
is therefore also unity.
When we adopt the stellar rotation rate, $\tilde{\Omega}=0.2$,
the corresponding rotational time scale
\EQ
\tau_\Omega=\Omega_0^{-1}
\EN
is then five, and the rotation period is $2\pi/\tilde{\Omega}\approx30$.
The characteristic time scale for the dynamo is the turbulent
diffusive time \citep[e.g.,][]{Sti74},
\EQ
\tau_{\rm TD}=R^2/\etaT,
\EN
which is around $500$ in our models.
Another interesting time scale for our models is the mass loss time,
\EQ
\tau_{\rm massloss}
=M/\bra{\dot{M}}\approx M_0/\dot{M}_0.
\EN
In our models, $M_0\approx7000$ and $\dot{M}_0=1$, so
$\tau_{\rm massloss}\approx7000$.
It turns out that the spindown time is of a similar
order of magnitude.
It is given by
\EQ
\tau_{\rm spindown}=J_\ast/\bra{\dot{J}},
\EN
where $J_\ast=\int_\ast\meanrho\varpi^2\Omega\,dV$ is the angular
momentum of the stellar envelope, with $\varpi=r\sin\theta$ being
the cylindrical radius, $\Omega=\meanU_\phi/\varpi$ is the local
angular velocity, and $\dot{J}$ is the angular momentum loss,
which we calculate in \Sec{AMflux}.
The asterisk on the integral denotes the volume of the envelope.
The mass loss and spindown times are the longest among the time scales
considered here, so the mass in the envelope cannot change significantly
during the time scales of interest for the wind and the dynamo.

\begin{figure}[t!]\begin{center}
\includegraphics[width=\columnwidth]{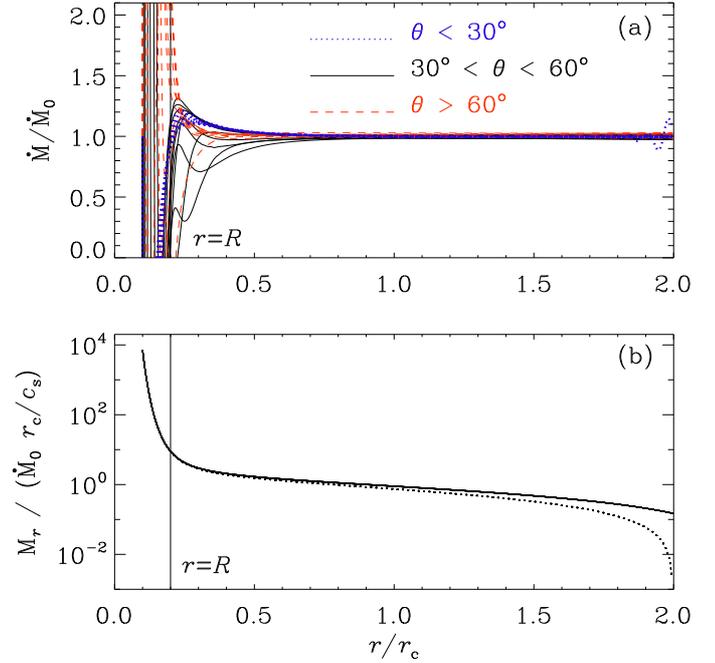}
\end{center}\caption[]{
Radial dependence of (a) $\dot{M}$ for different latitude ranges,
and (b) $M_r$ (solid line) for Model~A.
The dotted line in (b) refers to the mass within the computational
domain only, so it vanishes on $r=r_{\rm out}$.
}\label{ppmdot_mass_M4096a2_Q001_Om02}\end{figure}

\section{Results}
\label{Results}

After some preliminary studies at low resolution of $128\times32$
meshpoints with $\nuT=\etaT=10^{-2}\rcrit\cs$, we performed
high-resolution simulations with $4096\times1024$ meshpoints, where we
were able to decrease $\nuT$ and $\etaT$ to $8\times10^{-5}\rcrit\cs$.
These values are still above the physically motivated value, but for
numerical stability reasons, they cannot be decreased further without
invoking artificial viscosity and magnetic diffusivity.

Our main model is called Model~A, which has the solar value of $\Omega$
and a minimal amount of viscosity and magnetic diffusivity that can still
be tolerated.
Later, we also consider more rapidly rotating models cases (Models~B and C).

\begin{figure}[t!]\begin{center}
\includegraphics[width=\columnwidth]{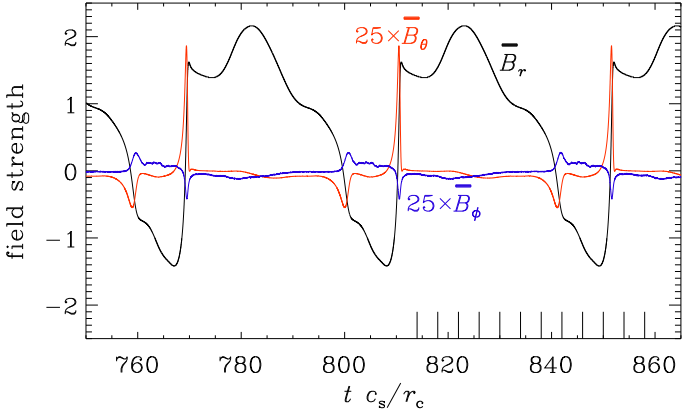}
\end{center}\caption[]{
Time series of the three magnetic field components
at one point for Model~A.
Here, $\meanB_\theta$ and $\meanB_\phi$ are multiplied by 25 to make
those components better visible.
We note that all three components of $\meanBB$ are asymmetric about zero.
The 12 long tick marks on the lower abscissa show the times for which
snapshots will be discussed later on.
}\label{ptime_M4096a2_Q001_Om02}\end{figure}

\subsection{Mass loss}

In \Fig{ppmdot_mass_M4096a2_Q001_Om02}a, we show the local mass loss density,
\EQ
\dot{M}(r,\theta,t)=4\pi r^2 \meanrho(r,\theta,t) \, \meanU_r(r,\theta,t),
\EN
whose average over $\theta$ and $t$,
$\bra{\dot{M}}=\int_0^\pi\int_{t_0}^{t_0+T}\dot{M}\,dt\,\sin\theta\,d\theta$,
is close to the initial value $\dot{M}_0$.
This is not too surprising, but it should be emphasized that this is not
enforced as a condition.
The good agreement suggests that the open boundary condition at the
bottom draws in a similar amount of mass at the inner boundary as what
is lost at the outer boundary.

\begin{figure*}[t!]\begin{center}
\includegraphics[width=\textwidth]{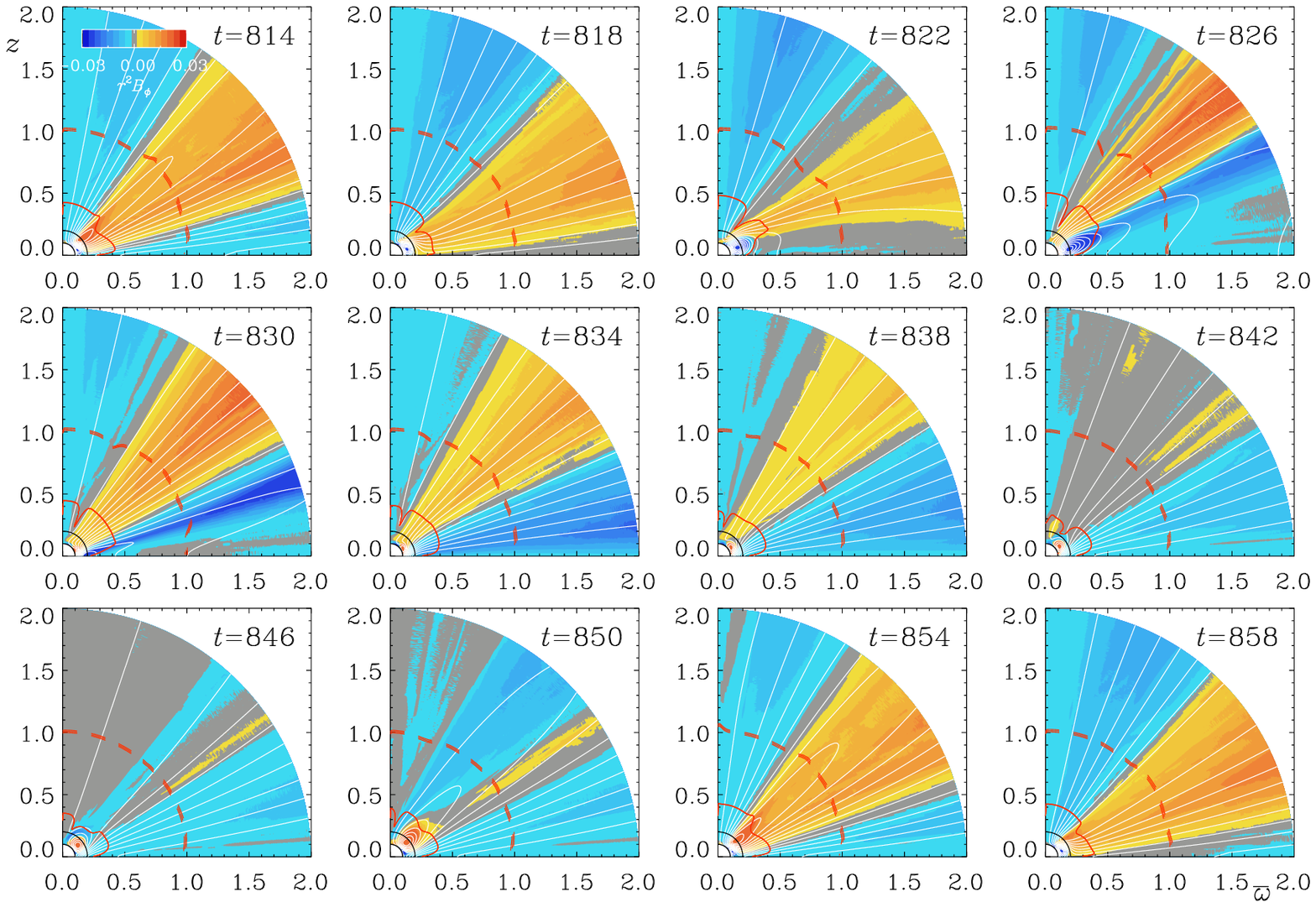}
\end{center}\caption[]{
Color representation of $r^2B_\phi(r,\theta)$ for different times for Model~A.
The nearly concentric red solid lines show the surfaces where $\meanU_r$
is transalfv\'enic and the red dashed ones show the surfaces where it is
transmagnetosonic.
The times correspond to the long tick marks of \Fig{ptime_M4096a2_Q001_Om02}.
}\label{ppvar_bb_panels}\end{figure*}

To get a sense of the radial mass distribution in our model, we plot
in \Fig{ppmdot_mass_M4096a2_Q001_Om02}b the cumulative mass,
\EQ
M_r(r,\theta,t)=\int_r^\infty 4\pi r'^2 \meanrho(r',\theta,t)\, dr',
\EN
for different values of $\theta$ at $t=858$.
We see that the total mass at $r=r_{\rm in}$ is about 7000 mass units;
one mass unit here is $\dot{M}_0 \rcrit/\cs$.
The mass above the surface is about 10, so 99.9\% of the total mass
in the computational domain is contained in the stellar envelope in
$r_{\rm in}\leq r\leq R$.
Thus, if no mass was replenished on the inner boundary, the time
it would take to lose all mass at the initial rate would be
$\tau_{\rm massloss}=M/\dot{M}=7000$.

\begin{figure*}[t!]\begin{center}
\includegraphics[width=\textwidth]{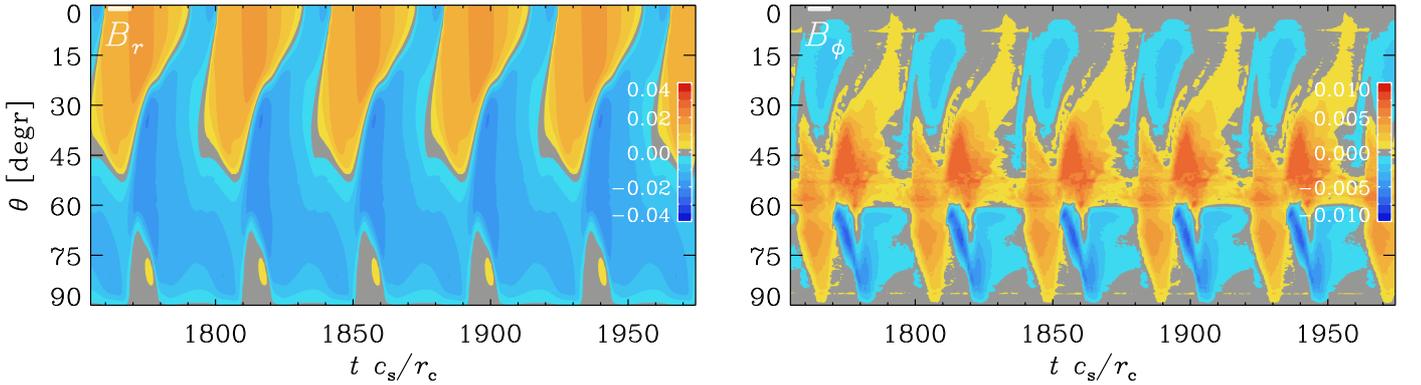}
\end{center}\caption[]{
Butterfly diagrams of $B_r(r,\theta)$ and $B_\phi(r,\theta)$ for Model~A
at $r/\rcrit=1.9$.
Again the asymmetry of those components with respect to zero,
which is different from the properties of the solar magnetic field.
}\label{pbutter_M4096a2_Q001_Om02_cont}\end{figure*}

\begin{figure*}[t!]\begin{center}
\includegraphics[width=\textwidth]{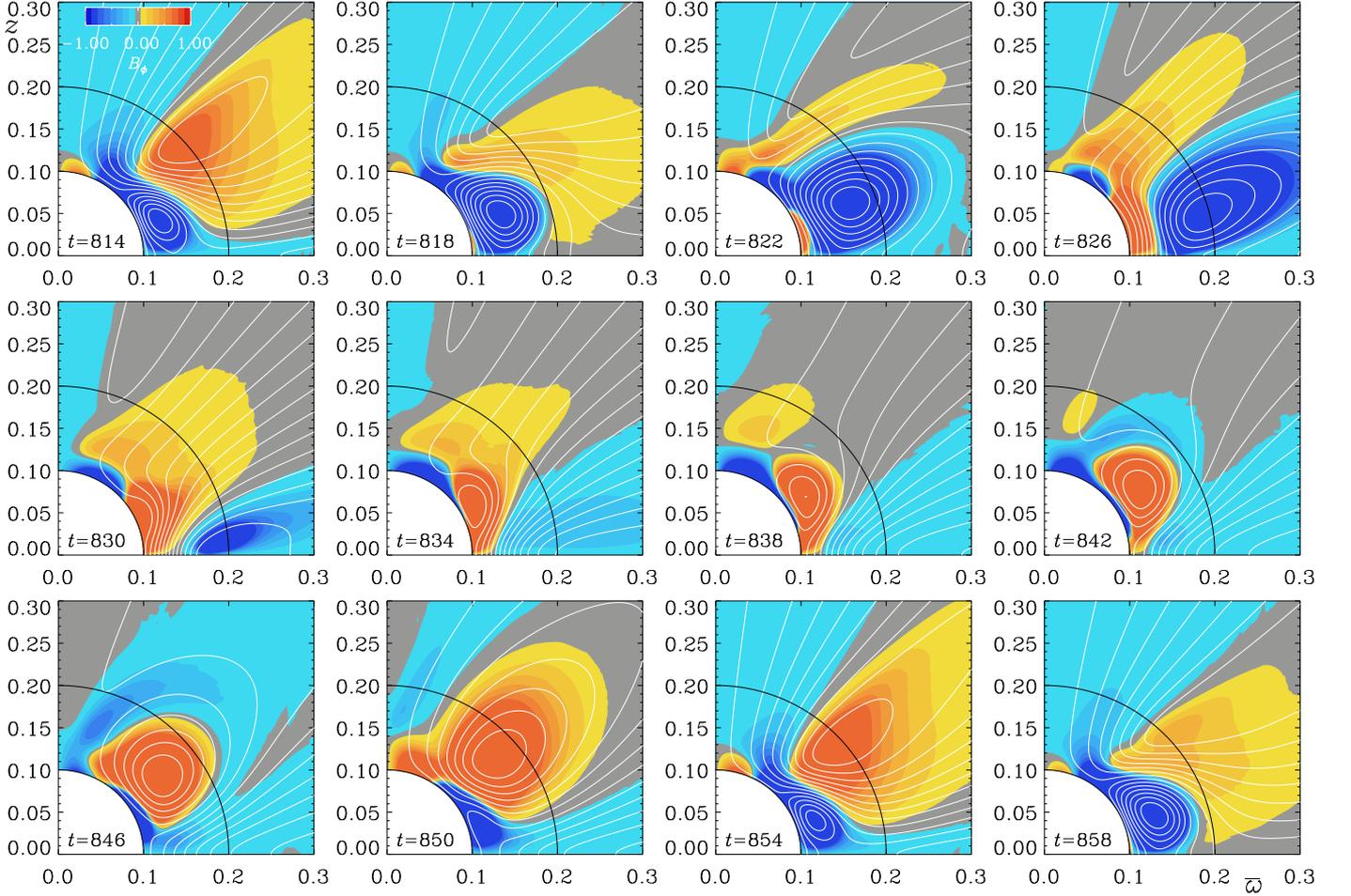}
\end{center}\caption[]{
Similar to \Fig{ppvar_bb_panels}, but this time with a color representation
of $B_\phi(r,\theta)$ showing only the region close to the center.
Note the occurrence of {\sf V}-shaped field lines during certain times
at $822\leq t\leq834$, and $846$.
The field shows radial outward migration during certain times: negative
$B_\phi$ at low latitudes for $814\leq t\leq826$, and positive $B_\phi$
at midlatitudes for $834\leq t\leq854$.
}\label{ppvar_bb_panels_inner}\end{figure*}

\begin{figure*}[t!]\begin{center}
\includegraphics[width=\textwidth]{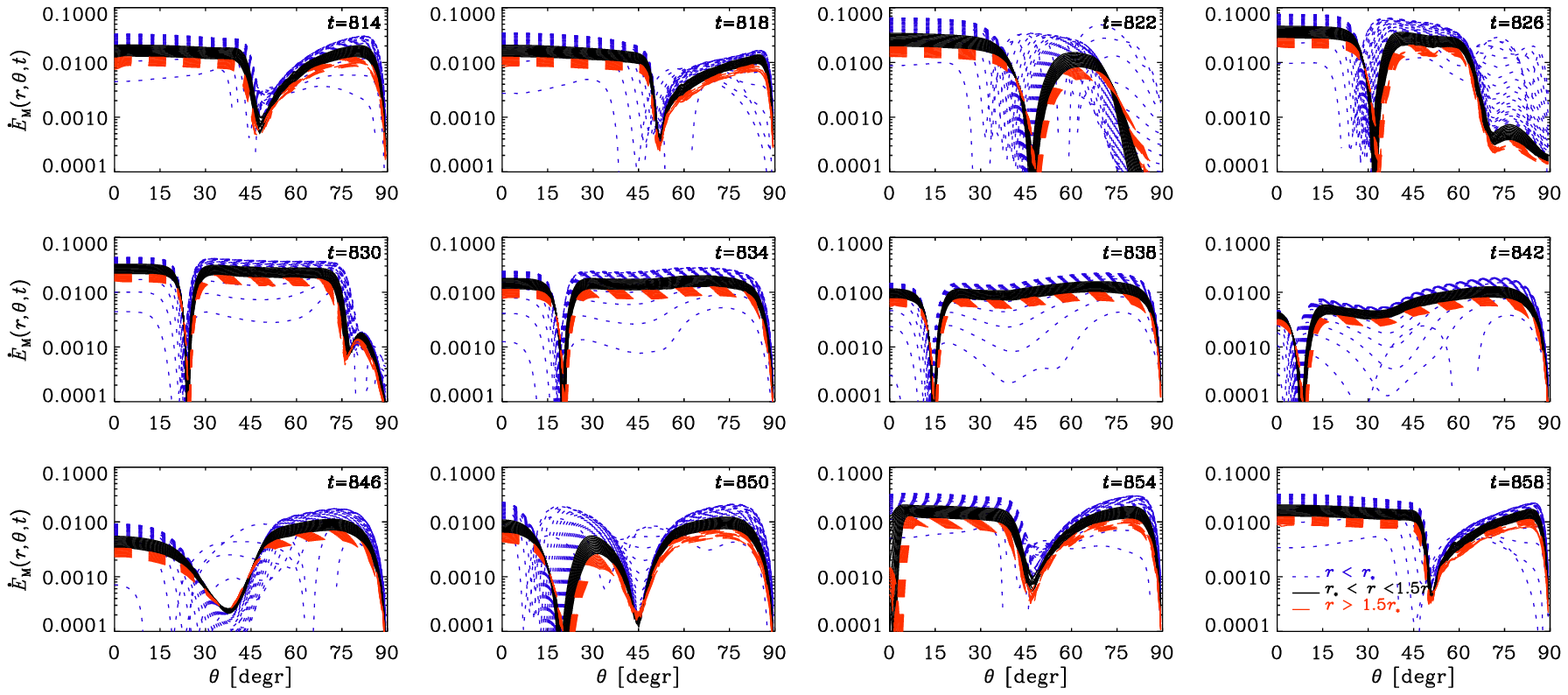}
\end{center}\caption[]{
Latitudinal dependence of the magnetic energy loss
at different times for Model~A.
Note the occurrence of a plateau for small values of $\theta$ for
$814\leq t\leq830$ and after $t=854$.
}\label{pb2_lat_panels_M4096a2_Q001_Om02}\end{figure*}

We emphasize at this point that the full stellar mass is undetermined,
because the value of Newton's constant $G$ never enters on its own.
We could, in principle, constrain it by assuming, for example, that the
density in the core is constant and equal to that at $r=r_{\rm in}$.
This would give for the minimal core mass $M_{\rm core}\gg36000$,
which is five times the mass in the envelope.
Using $GM_{\rm core}=2$, we find $G\ll6\times10^{-5}\cs^3/\dot{M}_0$,
which is satisfied by a large margin for the values quoted above.
We stress, however, that this estimate was done only for
illustrative purposes.

\begin{figure*}[t!]\begin{center}
\includegraphics[width=\textwidth]{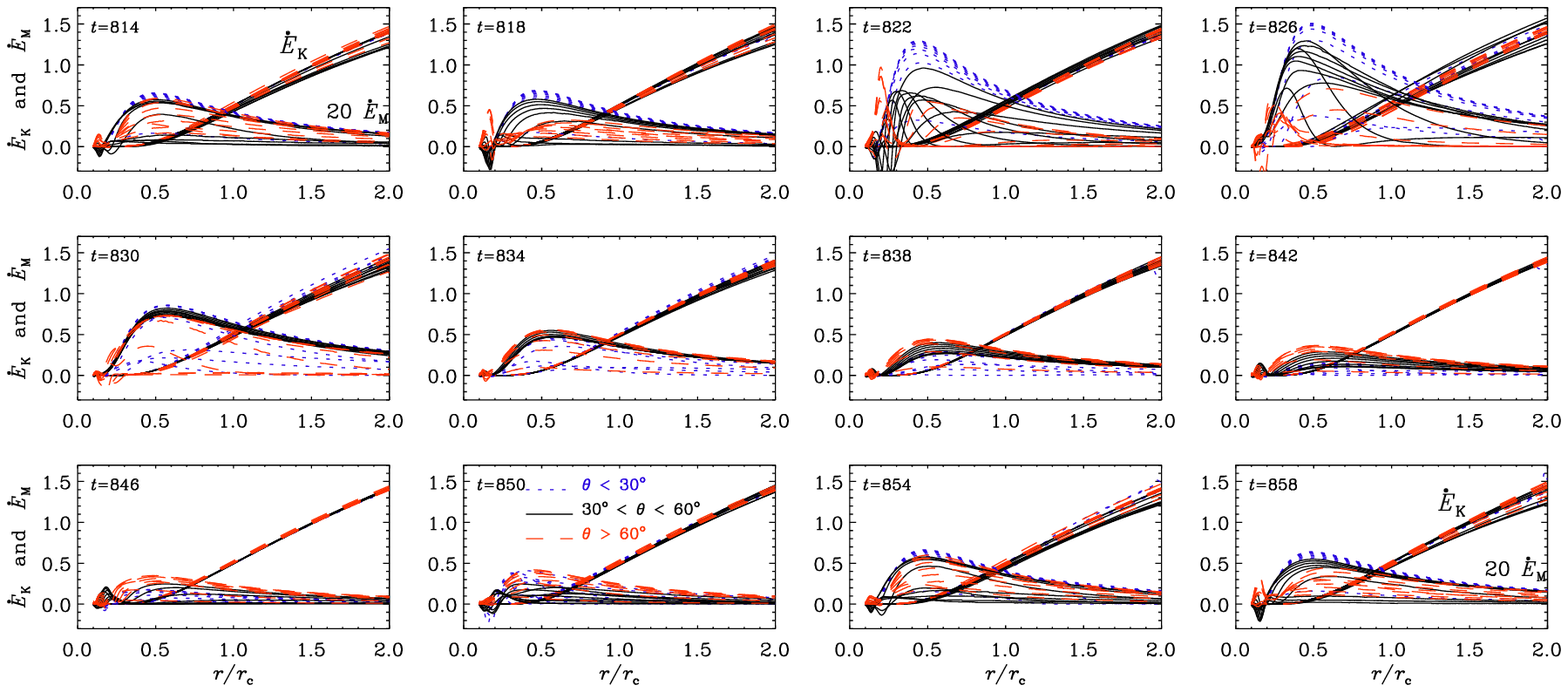}
\end{center}\caption[]{
Radial dependence of $\dot{E}_{\rm K}$ and $\dot{E}_{\rm M}$ for
different latitude ranges at different times for Model~A.
Note that $\dot{E}_{\rm M}$ has been multiplied by a factor of 20.
$\dot{E}_{\rm K}$ shows only little variability and always increases
radially outward, while $\dot{E}_{\rm M}$ has a maximum
near the Alfv\'en surface at $r/\rcrit\approx0.4$.
The maxima are particularly high for $822\leq t\leq826$.
}\label{ppLM_panels_M4096a2_Q001_Om02}\end{figure*}

\begin{figure*}[t!]\begin{center}
\includegraphics[width=\textwidth]{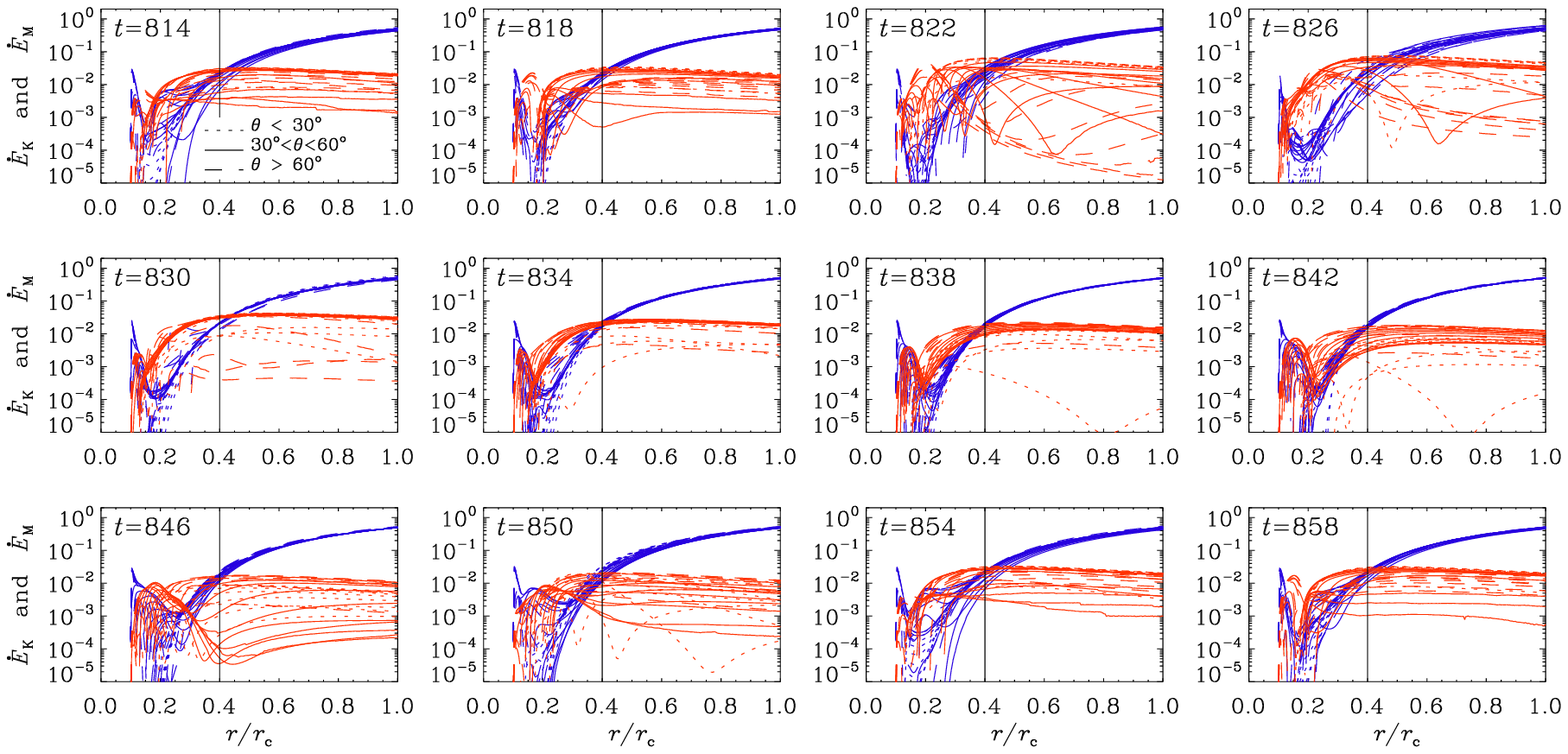}
\end{center}\caption[]{
Similar to \Fig{ppLM_panels_M4096a2_Q001_Om02}, for a semilogarithmic
representation, without having rescaled $\dot{E}_{\rm M}$.
Blue (red) lines indicate kinetic (magnetic) energy losses.
Note that $\dot{E}_{\rm M}\approx\dot{E}_{\rm K}$ near the Alfv\'en surface
at $r/\rcrit\approx0.4$, which is marked by a vertical line.
}\label{ppLM_log_panels_M4096a2_Q001_Om02}\end{figure*}

\subsection{Oscillatory model at solar rotation rate}

We focus on a simulation with the solar value of the angular velocity,
that is, $\tilde{\Omega}=0.2$ (Model~A).
In this case, the magnetic field is oscillatory, but in a rather
nonlinear fashion; see \Fig{ptime_M4096a2_Q001_Om02}, where we plot the
time dependence of the three magnetic field components at one point in
the wind.
The $\meanB_r$ component is positive most of the time and much smoother than
the $\meanB_\theta$ and $\meanB_\phi$ components.
The period $T$ is about 41 time units.
This corresponds to about $0.1\yr$, which is short compared with the
actual solar 22 year cycle, but still about five times longer than the
cycle period in the model of \cite{PBRS20}.
Their parameters are otherwise comparable to ours:
$\etaT\approx4\times10^{14}\cm^2\s^{-1}$ in both models,
$\dot{M}=3\times10^{-14}M_\odot\yr^{-1}$ (a third of our value),
an Alfv\'en radius of about two stellar radii,
and a domain size of 20 solar radii (twice our value).

\subsection{Magnetic field geometry}

In \Fig{ppvar_bb_panels} we show a sequence of magnetic field visualizations
at different times.
To make the magnetic field in the outer parts better visible, we multiply
$B_\phi$ by $r^2$.
Here, we show the time span from $\tilde{t}=814$ to $858$, covering just
a little over a period.
We overplot the surfaces where $\meanU_r$ is transalfv\'enic (solid
white lines), that is, where $\meanU_r$ exceeds the Alfv\'en speed
$\vA=(\meanBB^2/\mu_0\meanrho)^{1/2}$.
The surface is corrugated, but its mean radius is around $0.4\,\rcrit$.
We also shows the surfaces where $\meanU_r$ is transmagnetosonic,
that is, where $\meanU_r$ exceeds the fast magnetosonic speed
$c_{\rm ms}$ (dashed white line), which obeys $c_{\rm ms}^2=\cs^2+\vA^2$.
The mean radius of the magnetosonic surface is close to $\rcrit$.

Butterfly diagrams of $\meanB_r(\theta,t)$ and $\meanB_\phi(\theta,t)$
are shown in \Fig{pbutter_M4096a2_Q001_Om02_cont}.
The field in the wind does not show any migration in latitude, as is
expected from models of the solar dynamo.
\FFig{ppvar_bb_panels_inner} shows only the inner part of the domain.
We see regions with open and closed field lines at different times.
However, there is no clear magnetic field migration that manifests itself
in the Sun in a Maunder's butterfly diagram of sunspot locations versus
time and latitude.

It is interesting to note the appearance of {\sf V}-shaped field
lines in the panels for $t=822$--$834$ and perhaps also for $t=846$.
This means that there are magnetic field lines in the wind that are
not anchored in the star.
This may be a bit surprising, but we have to remember that the magnetic
field is time-dependent and the  medium electrically conducting.
The time-varying magnetic field can therefore induce toroidal currents
in the stellar wind, which then produce poloidal field lines that are
closed outside the star.
This phenomenon may be similar to what is known as ``switchbacks''
in the solar wind \citep{Bale,SCM20}.

\subsection{Poynting flux}

The wind carries with it not only mass, but also kinetic and magnetic energies.
The latter is quantified by the mean Poynting flux,
\EQ
F_{\rm Poy}(r,t)=\oint (\meanEE\times\meanBB/\mu_0)\,\cdot d\SSS,
\EN
where $\meanEE=\etaT\mu_0\meanJJ-\alpha\meanBB-\meanUU\times\meanBB$
is the mean electric field.
The magnetic energy loss is then $\dot{E}_{\rm M}=4\pi r^2 F_{\rm Poy}$.
In the steady state, $\bra{\dot{E}_{\rm M}}$ would be independent of $r$ if
there was no Ohmic dissipation and no conversion between kinetic and
magnetic energies in the wind.

As a good estimate for the magnetic energy loss of the solar wind,
\cite{BSBG11} computed
$\dot{E}_{\rm M}(r)\approx4\pi r^2 \bra{(\meanBB^2/2\mu_0)\,\meanU_r}$,
which they found to be on the order of $10^{18}\W$ and slowly
decreasing with radius.
Estimating the total magnetic energy content within the convection zone
based on a mean field of $300\G$ over the convection zone of volume
$4\pi(R^3-r_{\rm in}^3)/3$, we find a time scale of about 10 years,
which is comparable with the solar cycle period.

\FFig{pb2_lat_panels_M4096a2_Q001_Om02} shows the latitudinal dependence
of $\dot{E}_{\rm M}$ at different times for Model~A.
It depends not only on latitude and time, but also somewhat on radius.
There is a window at high latitudes where it is almost constant in $\theta$,
but the width of this window changes with time.
It can have a width of over $45\degr$ (e.g., at $t=818$ and 858), but it can
also be almost nonexistent (e.g., at $t=842$).
Comparing with \Fig{ppvar_bb_panels}, we see that this window 
of nearly constant $\dot{E}_{\rm M}$ corresponds to regions where the
radial field in the wind ist mostly negative.
The dips in $\dot{E}_{\rm M}$ correspond to regions where the radial
field is weak and changes sign.
Near the equator, $\dot{E}_{\rm M}$ shows a sharp drop for most times,
except for $t=826$.
Again, comparing with \Fig{ppvar_bb_panels}, we see that nothing special
happens near those dips, except that for $t=826$ the field is a bit weaker.
These dips are probably a consequence of the radial field reversal in the
equatorial plane and the existence of a field component that is purely
vertical to the equatorial plane, thus inhibiting the wind.

Next, we look at the radial dependence of the kinetic and magnetic
energy losses for different times and latitudes.
The result is shown in \Fig{ppLM_panels_M4096a2_Q001_Om02}, where we
define compute them as
\EQ
\dot{E}_{\rm K}=4\pi r^2(\meanrho\meanUU^2/2)\,u_r,
\EN
\EQ
\dot{E}_{\rm M}=4\pi r^2(\meanBB^2/2\mu_0)\,u_r,
\EN
respectively.
It turns out that $\dot{E}_{\rm M}$ is much smaller than $\dot{E}_{\rm K}$.
To accommodate both quantities in the same plot, we have multiplied
$\dot{E}_{\rm M}$ by a factor of 20.

We see that $\dot{E}_{\rm K}$ increases with radius.
This is a peculiar feature of isothermal models which is absent both
in isentropic models with constant specific entropy and in nonisentropic
models with variable specific entropy; see Figs.~9.18 and 9.20 of
\cite{Bra03}, respectively.
This is mainly because in those models the sound speed decreases with
radius in such a way that the Mach number still increases, just as in
the isothermal models.
Thus, the basic dynamics is similar in that the flow becomes supersonic.
In isothermal models, where the sound speed is constant, this transition
must always be accompanied by a radial increase of the wind speed.
In this sense, a polytropic model would seem more realistic, but it would
still ignore the internal energy or entropy equation, which would be even
more important for making our models more realistic, as is discussed below;
see \Sec{PlasmaBetas}.

\begin{figure*}[t!]\begin{center}
\includegraphics[width=\textwidth]{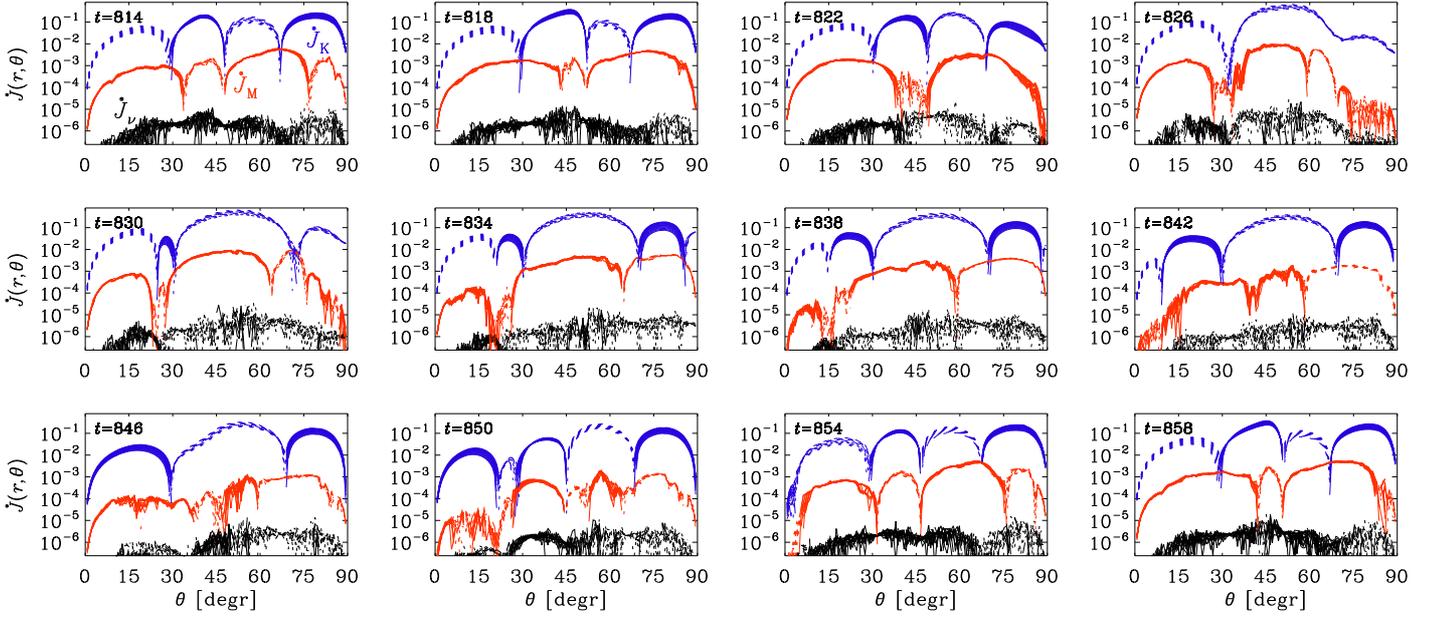}
\end{center}\caption[]{
Latitudinal dependence of the angular momentum loss $\dot{J}(r,\theta,t)$
for $1.5\leq r/r_{\rm c}\leq2$ at different times for Model~A.
The blue (red) lines refer to kinetic (magnetic) contributions,
and the black lines denote the turbulent viscous contribution.
Positive (negative) values are shown as solid (dotted) lines.
Note the strong latitudinal variability of $\dot{J}$.
}\label{pom_lat_panels_M4096a2_Q001_Om02}\end{figure*}

We also see that for $t=826$, when the field was a bit weaker further
out in the wind (\Fig{ppvar_bb_panels}), the kinetic energy
loss is particularly strong around the Alfv\'en surface; see
\Fig{ppLM_panels_M4096a2_Q001_Om02}.
At other times, especially for $t\leq842\leq850$, 
the kinetic energy loss is generally much weaker.
Comparing again with \Fig{ppvar_bb_panels}, this corresponds
to times when the radial field near the equator is strong.

In \Fig{ppLM_log_panels_M4096a2_Q001_Om02}, we show $\dot{E}_{\rm M}$ and
$\dot{E}_{\rm K}$ for $r\leq \rcrit$ as a semilogarithmic representation.
We see that $\dot{E}_{\rm M}\approx\dot{E}_{\rm K}$ at $r/\rcrit\approx0.4$.
The radial profiles of $\dot{E}_{\rm K}$ are fairly independent of $\theta$
and $t$.
This is because the wind is rather powerful and not much affected by
rotation or magnetic fields, which are the main factors that provide
non-spherically symmetric contributions to the system.

It is interesting to note that $\dot{E}_{\rm M}(r)$ has a maximum at
$r\approx \rcrit/2$.
This radius is a certain distance above the stellar surface and still
below the critical point.
This radius coincides with the Alfv\'en radius; see \Fig{ppvar_bb_panels}.
This is the point where most of the star's magnetic energy has been
deposited into the wind.
In the Sun, we expect that this energy deposition occurs in the corona.
One may tentatively associate the location of the maximum of
$\dot{E}_{\rm M}(r)$ with some representation of the star's corona,
although it is unclear whether there is any relation to the real corona
of the Sun.

At large radii, $r\gg \rcrit$, the magnetic energy loss declines slowly
with radius.
Such a decline has also been seen for the solar wind \citep{BSBG11}.
In the Sun, it may be connected with the conversion of magnetic energy
into heat.

\begin{figure}[t!]\begin{center}
\includegraphics[width=\columnwidth]{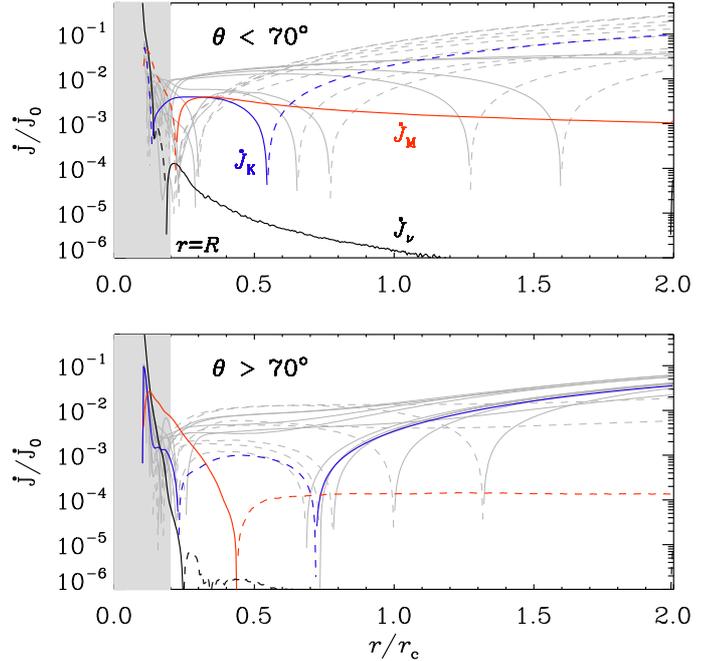}
\end{center}\caption[]{
Time averaged radial profiles of latitudinally averaged $\dot{J}$ for
$\theta<70\degr$ (upper panel) and $\theta>70\degr$ (lower panel).
The blue (red) lines refer to kinetic (magnetic) contributions,
and the black lines denote the turbulent viscous contribution.
Positive (negative) values are shown as solid (dotted) lines.
The kinetic contributions from different times are shown as gray lines.
The gray background on the left indicates the location of the stellar
envelope.
}\label{ppJdot_M4096a2_Q001_Om02}\end{figure}

\begin{figure*}[t!]\begin{center}
\includegraphics[width=\textwidth]{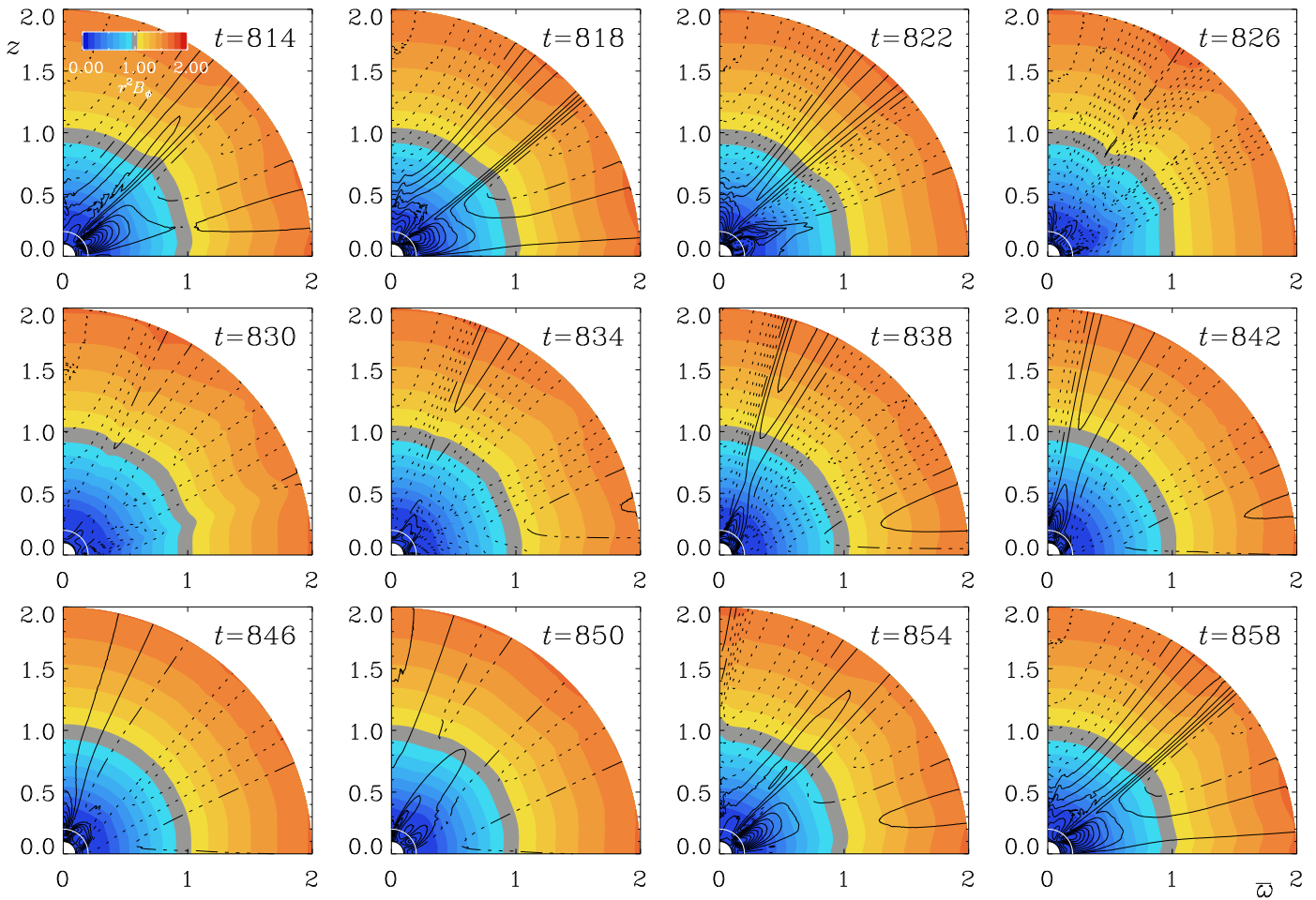}
\end{center}\caption[]{
Angular velocity contours superimposed on a color representation of
$\meanU_r(r,\theta)$ for Model~A.
Positive (negative) values of $\Omega$ are shown as solid (dotted)
black contours.
Note the extended region at midlatitudes were $\Omega<0$.
}\label{ppvar_uu_several}\end{figure*}

\subsection{Angular momentum flux}
\label{AMflux}

There are no sinks or sources to the angular momentum
density, $\meanrho\varpi^2\Omega$, and it therefore satisfies a
conservation equation of the form \citep{Mes68,Mes99}
\EQ
\frac{\partial}{\partial t}\left(\meanrho\varpi^2\Omega\right)=
-\nab\cdot\FF^{\rm AM},
\EN
where
\EQ
\FF^{\rm AM}=
\meanrho\varpi^2\Omega\meanUU-\varpi\meanB_\phi\meanBB/\mu_0
-\meanrho\nuT\varpi^2\nab\Omega
\EN
is the angular momentum flux.
Analogously to the energy loss, the expression for the angular
momentum loss is $\dot{J}=4\pi r^2 F_r^{\rm AM}$, which is shown in
\Fig{pom_lat_panels_M4096a2_Q001_Om02} for $1.5\leq r/r_{\rm c}\leq2$
for the kinetic, magnetic, and viscous contributions, $\dot{J}_{\rm K}$,
$\dot{J}_{\rm M}$, and $\dot{J}_\nu$, respectively.
We see that the angular momentum flux is highly structured, with positive
and negative contributions at different latitudes and times.
At these radii,
the kinetic term proportional to $\meanU_\phi\meanU_r$ dominates over the
magnetic term proportional to $\meanB_\phi\meanB_r$, and the turbulent
viscous term is negligible.

The strongly negative contributions to the angular momentum flux are
unexpected and may be connected with the time dependence of the solution.
It may be of interest to study angular momentum fluxes along magnetic field
lines; see the work of \cite{PM17}, who compare flow speeds along different
field lines.
For our unsteady wind solutions, this procedure may no longer be
particularly advantageous.
However, to get some idea about the latitudes contributing to the
negative angular momentum flux, we show in
\Fig{ppJdot_M4096a2_Q001_Om02} the radial dependence of the
time- and latitude-averaged profiles of $\dot{J}$ separately for the
cones $\theta<70\degr$ (away from the equator) and $\theta\ge70\degr$
(around the equator).
We see that negative angular momentum fluxes dominate and
originate mainly from regions away from the equator.
Nevertheless, in the range $0.2\leq r/r_{\rm c}\leq0.6$, 
$\dot{J}_{\rm K}$ and $\dot{J}_{\rm M}$ can be of comparable
magnitude, as is expected from the theory of \cite{WD67}.
This range agrees well with the Alfv\'en radius; see
\Fig{ppvar_bb_panels}.

To understand the variability of $\Omega$ and the occurrence of
negative values at certain times, we show in \Fig{ppvar_uu_several}
angular velocity contours superimposed on a color representation of
$\meanU_r(r,\theta)$.
Interestingly, $\Omega$ is often negative over an extended range of
mid-latitudes.
As we have seen above, this is chiefly responsible for the inward
angular momentum transport discussed above.
This could be related to our rather primitive modeling of the
hydrodynamics inside the star, which lacks realistic differential
rotation, for example.
We return to this question briefly in the conclusions.
We also note that $\meanU_r(r,\theta)$ shows clear latitudinal variations.
The occurrence of regions with negative angular momentum transport is
interesting in view of the recent discovery of fast wind episodes observed
with Parker Solar Probe at certain longitudes \citep{Finley20a,Finley20b}.
Our model is of course axisymmetric and cannot address longitudinal
variations, but it reminds us that negative angular momentum transport
is not impossible.

We should point out that $\dot{J}$ is given here in standard units
where $\dot{M}=\cs=1$.
Therefore, \Fig{pom_lat_panels_M4096a2_Q001_Om02} can be directly
interpreted as a plot of the mean-field (MF) analogue of the
\cite{SS73} parameter,
\EQ
\alpha_{\rm SS}^{\rm MF}
=(\meanrho\meanU_r\meanU_\phi-\meanB_r\meanB_\phi/\mu_0)/\cs^2.
\EN
Here, the superscript MF indicates that this expression is
applied to the two-dimensional mean fields rather than to the
fluctuations, as in the usual turbulent case.
This parameter is also frequently used in solar wind studies
\citep[see Eq.~(2) of][]{Finley}; see also \cite{KG99}, \cite{Rev15},
and \cite{PM17} for earlier two-dimensional stellar wind models.

The angular momentum in the dynamo zone is $J_\ast\approx68$ in our units.
Owing to cancelation, it is difficult to determine reliable values
of $\dot{J}$ and $\alpha_{\rm SS}^{\rm MF}$, but for the purpose of a
preliminary assessment, it suffices to estimate $\dot{J}\approx0.01$.
As we discuss below in more detail, there can be certain periods
where $\dot{J}$ can even be negative.
This then implies spindown or spinup at a rate $\tau_{\rm spin}\approx7000$,
which is indeed similar to the value of $\tau_{\rm massloss}$ quoted
in \Sec{TimeScales}.
It may well be that $\alpha_{\rm SS}^{\rm MF}$ is much less than 0.01.
This would then imply an even larger value of $\tau_{\rm spindown}$.

\subsection{Resulting dynamo parameters}

In our model, differential rotation is automatically established as a
result of magnetic braking.
Since our turbulent viscosity is assumed to be purely isotropic,
differential rotation can only result from the torque on the star
established by the magnetized wind \citep{Mes68}.
This leads to a nearly constant angular momentum per unit mass,
that is, $\varpi^2\Omega\approx\const$.
The contours of constant angular velocity tend to approach a pattern
that is close to cylindrical, as will be discussed below in the context
of rapid rotation.
Given that $\Omega\propto\varpi^{-2}$, the angular
velocity difference between the $r_{\rm in}$ and $R$ is
$\Delta\Omega=(1-r_{\rm in}^2/R^2)\Omega_0=0.75\,\Omega_0$.
Therefore, we have for the second dynamo parameter in \Eq{CalpCom} the
values $C_\Omega=75$, $375$, and $3750$ for $\tilde{\Omega}=0.2$, $1$,
and $10$, respectively.
The first dynamo parameter in \Eq{CalpCom} is
$C_\alpha=125$, where we have used
$\tilde{\alpha}_0\equiv\alpha_0/\cs=0.05$
for Model~A, and $\etaT=8\times10^{-5}\rcrit\cs$.

\subsection{Rapid rotation}

The study of models at rapid rotation is motivated by the interest in
understanding the evolution of magnetic activity of young stars, that is,
before they have slowed down to the solar rotation rate.
For us, there is also another motivation in that all our models were
of $\alpha^2$ type, that is, the $\Omega$ effect was weak and
$C_\Omega$ was not much larger than $C_\alpha$, as required for
an $\alpha\Omega$ dynamo \citep{BS05}.
To increase $C_\Omega$, the rotation rate could be increased.
Another possibility is to lower $C_\alpha$.
However, to prevent the dynamo from decaying, one would need to decrease $\etaT$
even further, but this is computationally difficult.

For rapid rotation, the magnetic field lines and contours of the
toroidal magnetic field are much more concentrated to the bottom of the
dynamo region, $r\approx r_{\rm in}$.
At faster rotation, the contours become more cylindrical.
This is an effect of the Taylor--Proudman theorem and results generally
in small variations along the rotation axis.

\begin{figure}[t!]\begin{center}
\includegraphics[width=.49\columnwidth]{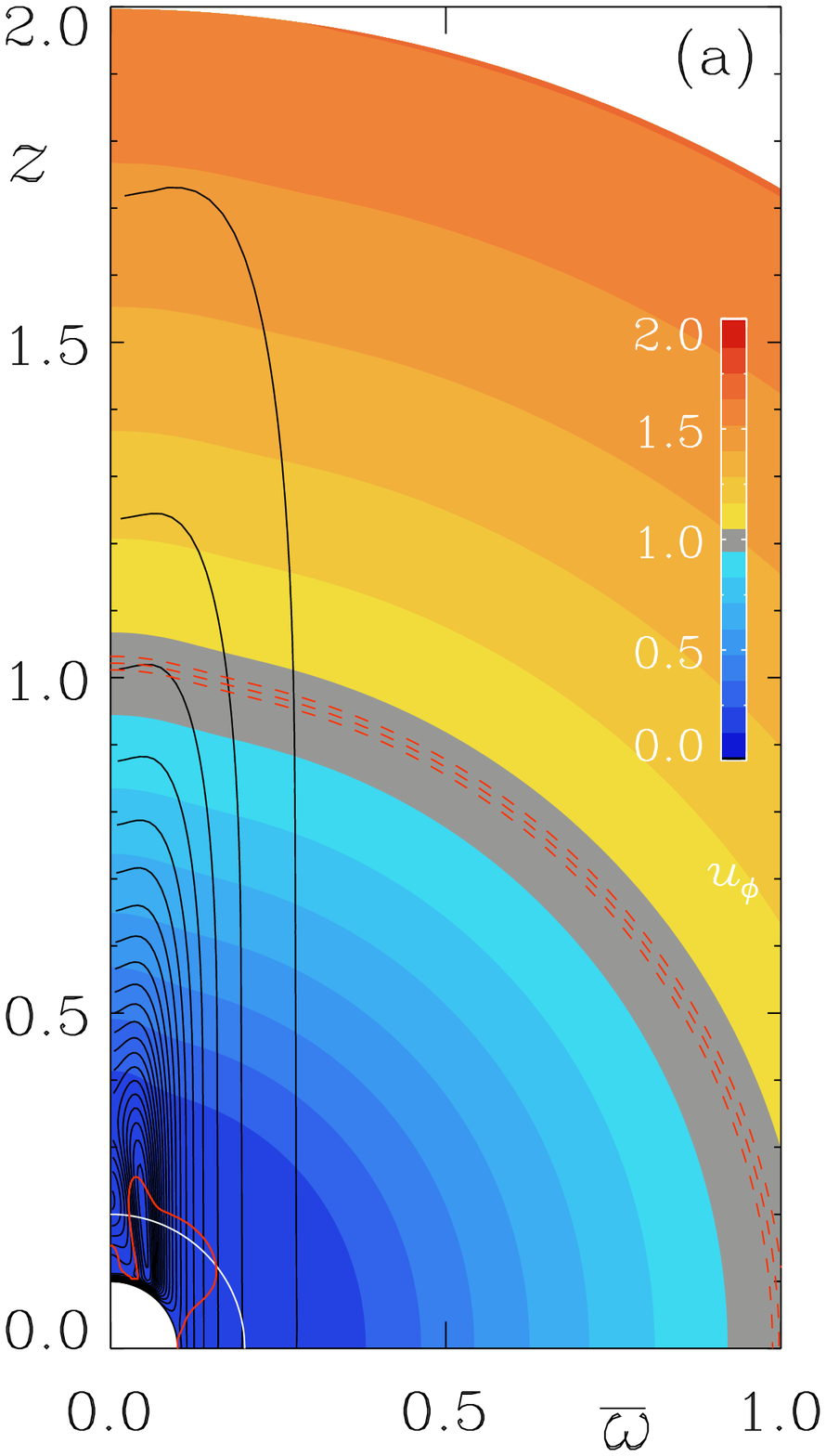}
\includegraphics[width=.49\columnwidth]{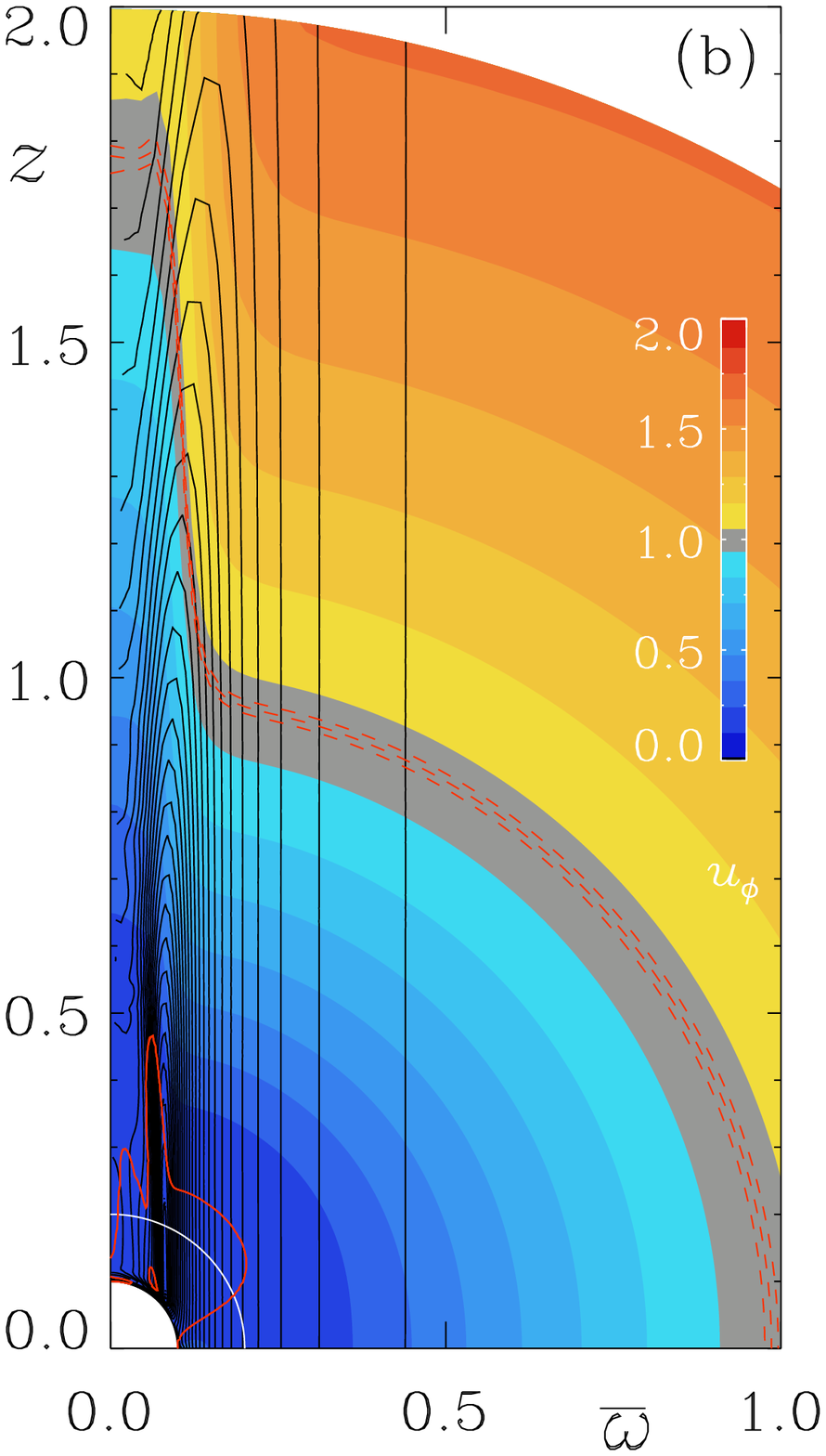}
\end{center}\caption[]{
Angular velocity contours superimposed on a color representation of
$\meanU_r(r,\theta)$ for Model~B (a) with $\tilde{\Omega}=1$
and Model~C (b) with $\tilde{\Omega}=10$.
The nearly concentric red solid lines show the surfaces where $\meanU_r$
is transalfv\'enic and the red dashed ones show the surfaces where it is
transmagnetosonic.
}\label{ppvar_uu}\end{figure}

\begin{figure}[t!]\begin{center}
\includegraphics[width=\columnwidth]{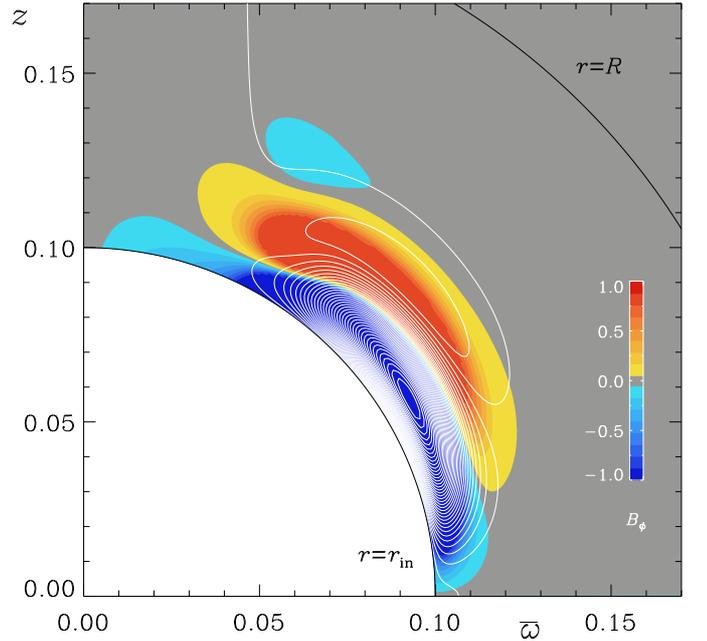}
\end{center}\caption[]{
Magnetic field lines superimposed on a color representation of
$\meanB_\phi(r,\theta)$ for Model~B with $\tilde{\Omega}=1$.
Strong fields only occur near $r=r_{\rm in}$; the weak-field regions
elsewhere cannot be seen.
}\label{ppvar_bb_M4096b1_Q001_Om1}\end{figure}

\begin{figure}[t!]\begin{center}
\includegraphics[width=\columnwidth]{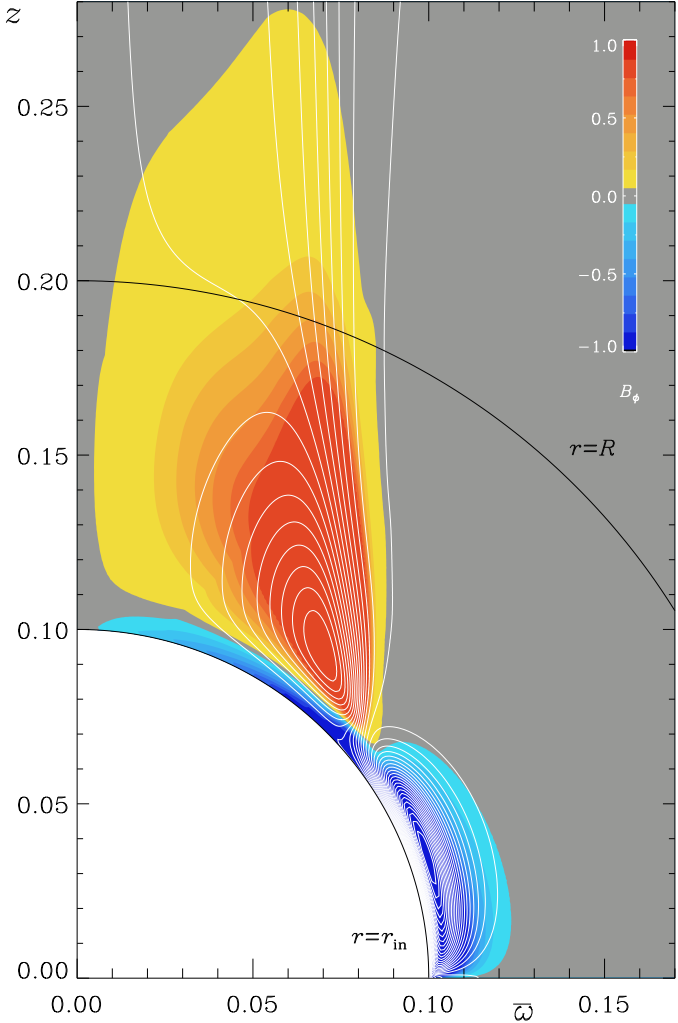}
\end{center}\caption[]{
Similar to \Fig{ppvar_bb_M4096b1_Q001_Om1}, but for Model~C with
$\tilde{\Omega}=10$.
Stronger fields now extend along the axis outside the star.
}\label{ppvar_bb_M4096a13_Q01_Om10}\end{figure}

The Taylor--Proudman theorem applies primarily to the angular velocity contours.
This can be seen by writing the relevant part of the $\meanUU\cdot\nab\meanUU$
nonlinearity of \Eq{dUmean} in the form
\EQ
\pphi\cdot\nab\times\left(-\meanUU\cdot\nab\meanUU\right)_{\rm p}
=\varpi\frac{\partial}{\partial z}\Omega^2 + ...,
\EN
where $\Omega=\meanU_\phi/\varpi$ is the local angular velocity,
and the dots indicate the presence of other terms not relevant here.
In \Fig{ppvar_uu}a we show contours of $\Omega$ together with a color-coded
representation of $\meanU_r$.
We see that the $\Omega$ contours are already strongly cylindrical for
$\tilde{\Omega}=1$.
As we increase the value of $\tilde{\Omega}$ to 10,
the cylindrical contours begin to extent much further out along the
rotation axis; see \Fig{ppvar_uu}b.

For $\tilde{\Omega}=10$, the radial velocity develops a marked indentation
inside of what is known as the inner tangent cylinder where
\EQ
\varpi\ge r_{\rm in}\quad\mbox{(inner tangent cylinder)};
\EN
see \Fig{ppvar_uu}b.
Here the outflow is suppressed and supersonic flows occur only for
$z\ge2\rcrit\approx r_{\rm out}$, that is, near the outer boundary of
the computational domain.
For $\tilde{\Omega}=1$, by comparison, the contours of
$\meanU_r(r,\theta)$ are almost perfectly spherically symmetric --
much more so than even for the case with $\tilde{\Omega}=0.2$; cf.\
\Fig{ppvar_uu_several}.
Similar results have also been found by \cite{WS93} in their rotating
models where a central dipole magnetic field was assumed.

\begin{table}[b!]\caption{
Summary of the simulations discussed in this paper.
}\vspace{12pt}\centerline{\begin{tabular}{crcccccc}
Model & $\tilde{\alpha}$ & $Q_\alpha$ & $\tilde{\Omega}$ &
$C_\alpha$ & $C_\Omega$ & $\meanB_{\max}$ & $P_{\rm cyc}$ \\
\hline
A &$ 0.05$& $10^{-2}$ &0.2 & 125 &   75 & 6--13& 41.0 \\
B &$ 0.1 $& $10^{-2}$ & 1  & 250 &  375 & 16.0 & ---  \\
C &$ 0.1 $& $10^{-1}$ & 10 & 250 & 3750 &  8.8 & ---  \\
\label{Tab2}\end{tabular}}\end{table}

It turns out that our models are now no longer oscillatory and are
thus still not of $\alpha\Omega$ type, contrary to what was originally
hoped for.
Visualizations of the toroidal and poloidal fields for Models~B and C
are shown in \Figs{ppvar_bb_M4096b1_Q001_Om1}{ppvar_bb_M4096a13_Q01_Om10},
respectively.
The fields are strong only inside the star, where the dynamo is active.
Outside the star, the field is much weaker and not visible in our
graphical representation, but it is never vanishing.

\begin{figure}[t!]\begin{center}
\includegraphics[width=\columnwidth]{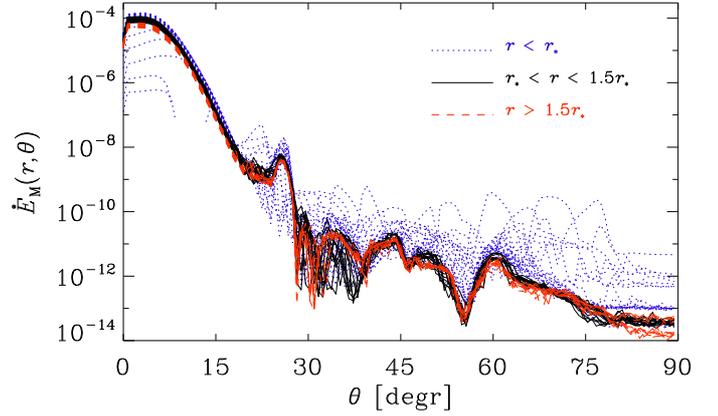}
\end{center}\caption[]{
Latitudinal dependence of $\dot{E}_{\rm M}$ for different radius
ranges for Model~B.
Note that $\dot{E}_{\rm M}$ is large only near the axis.
}\label{pb2_lat_M4096b1_Q001_Om1}\end{figure}

To discuss the nonoscillatory nature of these two models, it is
useful to consider the dynamo parameters $C_\alpha$ and $C_\Omega$.
We find $C_\alpha=125$ and $C_\Omega=3750$ for $\etaT=\nuT=8\times10^{-5}$;
see \Tab{Tab2}.
To get an idea about the latitudinal variation of the magnetic field in
the wind, the plot $\dot{E}_{\rm M}$ as a function of $\theta$ for different
radii.
The result is shown in \Fig{pb2_lat_M4096b1_Q001_Om1}.
It turns out that the magnetic activity is confined to a narrow cone
with an opening angle of about $15\degr$.

Noticeable magnetic energy losses are found only near the rotation axis.
As a function of radius, similarly to the case of slow rotation,
$\dot{E}_{\rm M}(r)$ has a maximum somewhere in $R<r<\rcrit$, which is
where the Alfv\'en point lies.
Furthermore, Model~B has a much smaller magnetic
energy loss at large radii than Model~A.

The model shows similarities with earlier simulations of
outflows emanating from stellar accretion disk dynamos \citep{vRBDS03,vRB04},
but there the opening angle was closer to $30\degr$.
In the present simulations, the opening angle is essentially zero.
It corresponds to a cylinder in which most of the magnetic fields are
ejected, although the flow speed here is strongly reduced.

\begin{figure}[t!]\begin{center}
\includegraphics[width=\columnwidth]{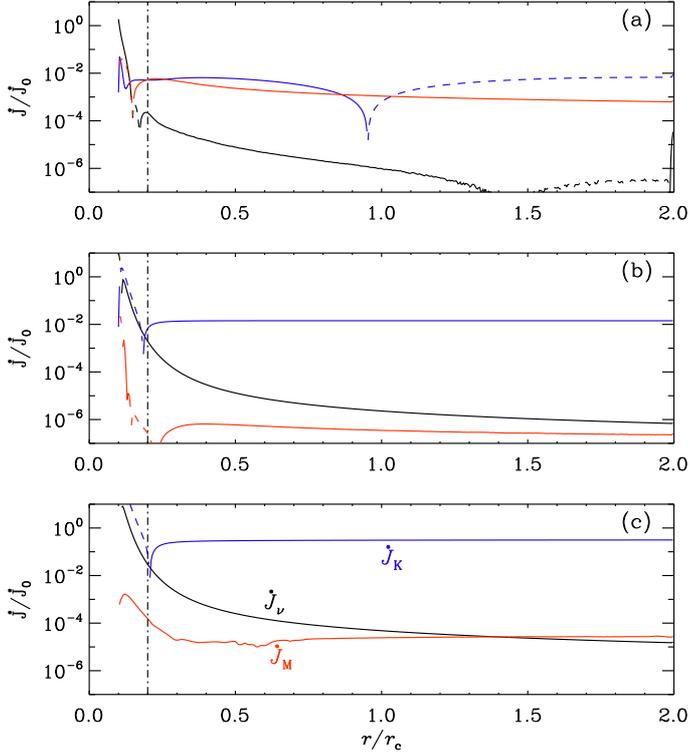}
\end{center}\caption[]{
Radial profiles of the latitudinally averaged $\dot{J}$ for
Models~A--C in panels (a)--(c).
The blue (red) lines refer to kinetic (magnetic) contributions,
and the black lines denote the turbulent viscous contribution.
Positive (negative) values are shown as solid (dotted) lines.
The total (kinetic, magnetic, and viscous) angular momentum transport
is dominated by the kinetic contribution, except for Model A, where the
magnetic contribution is rather strong, but negative in the outer parts.
}\label{ppJdot_single}\end{figure}

For these rapidly rotating models, we expect significant outward angular
momentum transport.
To demonstrate this in more detail, we show in \Fig{ppJdot_single} the
radial profiles of the latitudinally averaged $\dot{J}$ for Models~A--C
for the kinetic, magnetic, and viscous contributions,
just as we did in \Fig{ppJdot_M4096a2_Q001_Om02}.
Since Models~B and C are steady, time averaging is only needed for Model~A.

\Fig{ppJdot_single} shows that in Models~B and C, the angular
momentum transport is outward and $\dot{J}$ is independent of $r$
throughout most of the wind.
For Model~A, however, the time-averaged angular momentum transport
becomes negative some distance away from the Alfv\'en point.
Furthermore, $\dot{J}$ is more than ten times larger in Model~C than
in the ten times more slowly rotating Model~B.
For Model~B, the viscous contribution exceeds the magnetic one at all
radii, while in Model~C, the magnetic contribution exceeds the viscous
one for $r/r_{\rm c}>1.5$.
Inside the star, the angular momentum transport is negative and caused
by a strong poleward circulation.
The viscous contribution is also rather strong, but positive.

\begin{figure}[t!]\begin{center}
\includegraphics[width=\columnwidth]{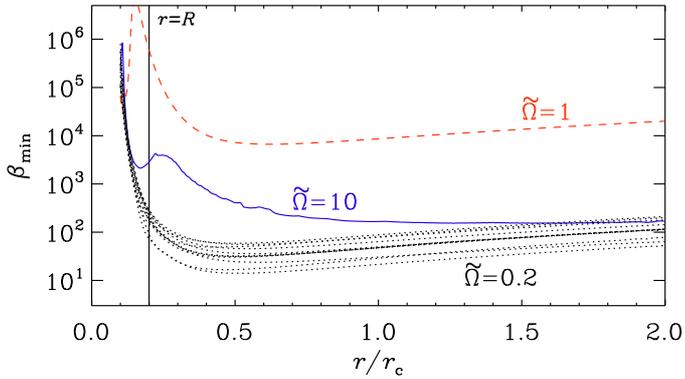}
\end{center}\caption[]{
Radial dependence of the plasma $\beta$ for Models~A (dotted black lines),
B (dashed red line), and C (solid blue line).
}\label{ppbeta}\end{figure}

\subsection{Comparison of the plasma betas for our models}
\label{PlasmaBetas}

We have seen that in Model~A with the slowest rotation, the angular
momentum flux was occasionally inward, especially at midlatitudes.
We then considered Models~B and C with faster rotation in the hope that
not only the outward angular momentum flux would be outward, but also
that the dynamo in the star would be in the $\alpha\Omega$ regime.
We found that the angular momentum flux was then indeed outward,
but the dynamo was still in the $\alpha^2$ regime.
In the introduction, we did already emphasize that the lack of a
cool photosphere just beneath the corona was ignored.
This makes it generally very difficult to reach low plasma betas,
which we define as
\EQ
\beta=2\rho\cs^2/\BB^2.
\EN
In \Fig{ppbeta}, we plot the radial dependencies of the minimum value
of $\beta$, $\beta_{\min}$, for Models~A--C.
We see that the largest values of $\beta_{\min}$ occur for Model~B
with an intermediate angular velocity.
Increasing the angular velocity further (Model~C) increases the
field strength and does therefore also lead to a smaller value of
$\beta_{\min}$.
The smallest values occur for Model~A.
This is mainly because Model~A is the only model where the magnetic
field in the wind is of comparable strength at all latitudes.
For faster rotation, the field in the wind is strongly concentrated
around the axis.

Let us now return to the potential role of the photosphere.
The photosphere of a star is the region where it cools and loses
specific entropy.
Everywhere else in the wind, the specific entropy does not change much,
and therefore the potential enthalpy must be approximately constant
\citep{vRBDS03}.
The potential enthalpy is defined as $H=h+\Phi$, where $h=\cp T$ is the
specific enthalpy with $\cp$ being the specific heat at constant pressure
and $T$ the temperature, and $\Phi=-GM/r$ is the potential energy.
Hydrostatic equilibrium requires that 
\EQ
0=-\nab H+h\nab s/\cp,
\EN
where $s$ is the specific entropy.
For the corona, this implies $T=GM/r\cp\approx2\times10^6\K$, which is
realistic and agrees also with our model.
Toward the photosphere, $T$ decreases abruptly because of surface cooling,
and therefore the density increases abruptly.
Thus, the density would then be much larger than what was possible in our models.
This, in turn, would allow us to reach much larger field strengths and
therefore smaller plasma betas.

\begin{figure}[t!]\begin{center}
\includegraphics[width=\columnwidth]{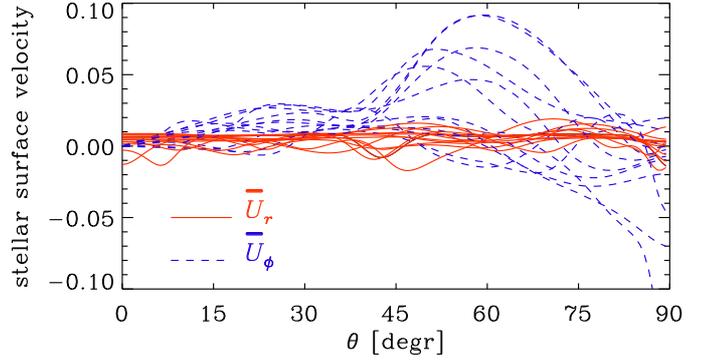}
\end{center}\caption[]{
$\meanU_r$ (red solid lines) and $\meanU_\phi$ (blue dashed lines) at
the stellar surface at $r=R$ as a function of colatitude for the same
times as in \Fig{ppJdot_M4096a2_Q001_Om02}.
$\meanU_r$ is usually positive, but $|\meanU_\phi|$ is much larger
and most of the time in the prograde direction, but sometimes it is
retrograde, which is a consequence of the low moment of inertia
of the stellar envelope in our model.
}\label{ppJdot_M4096a2_Q001_Om02_surf}\end{figure}

Another important consequence of having larger densities in the
stellar envelope would be that the angular velocity at the stellar
surface would always be in the prograde direction.
In our present models, this is not always the case, as can be seen
from \Fig{ppJdot_M4096a2_Q001_Om02_surf}, where we show the  radial
and azimuthal velocities at the stellar surface.
We see that the local rotational velocity is there occasionally in the
retrograde direction, especially near the equator.

\section{Conclusions}
\label{Conclusions}

Our work has shown that a simplified realization of a dynamo with a
stellar wind can easily be treated self-consistently in one and the same
model, provided certain compromises are being made.
The assumption of an isothermal equation of state has simplified
matters conceptionally.
Relaxing this restriction would allow us to include the energy deposition
in the corona and to model the effects of a sharp density drop at the
stellar surface.
This might require a significant increase in resolution near the surface,
which in turn requires the use of a nonuniform mesh.
Another restriction has been the use of a relatively large turbulent
magnetic diffusivity and viscosity.
This was mainly needed to resolve shocks that develop within the wind.
Those typically emerged in response to rapid changes in the magnetic field.
This could probably be avoided by allowing for an additional shock
viscosity, but this has been avoided in the present work.
On the other hand, the angular momentum flux associated with turbulent
viscosity was already negligible, so its presence may not have caused
any artifacts.

Future work might involve the inclusion of a $\Lambda$ effect
\citep{Rue80,Rue89}, which would allow for the development
of differential rotation in the stellar envelope.
Without including the effects of stellar winds, such models with combined
$\alpha$ and $\Lambda$ effects were studied by \cite{BMRT90,BMRT91}, who
found significant alignment of the $\Omega$ contours with the rotation
axis unless the baroclinic term was also included \citep{BMT92}.
But this may change when their boundary condition on $r=R$ is replaced
by a continuous transition to the solar exterior; see \cite{WKMB13}
for spherical convection simulations with a simplified representation
of a stellar corona.

The inclusion of the $\Lambda$ effect might allow us to model the stellar
dynamo more realistically.
It would be interesting to see how this affects the angular momentum
transport and whether it could help in producing predominantly outward
angular momentum transport in cases of slow rotation.
It might then allow us to study dynamos in the $\alpha\Omega$ regime.
This has not been possible in the present model for reasons that are
not entirely clear, because the value of $C_\Omega$ was thought to be
already large enough.
There could have been other side effects arising from the coupling to the
outflow that are not yet fully understood.
Nevertheless, it is interesting to note that the inward angular momentum
transport occurs even in the Sun within fast-wind regions at certain
longitudes; see \cite{Finley20a,Finley20b}.

Another important aspect requiring further attention is the study of
angular momentum losses from mean-field stresses.
Our work has shown that the angular momentum loss can be quantified in
terms of a nondimensional Shakura--Sunyaev parameter.
This is a somewhat unusual concept in the context of stellar winds,
but it may help putting the theories of turbulent stellar winds and
accretion disks on a common footing.

\begin{acknowledgements}
We thank the referee for many useful remarks and suggestions that have
significantly improved the manuscript.
This work was supported in part through the Erasmus+ Programme of
the European Union (P.J.) and the Swedish Research Council,
grant 2019-04234 (A.B.).
We acknowledge the allocation of computing resources provided by the
Swedish National Allocations Committee at the Center for
Parallel Computers at the Royal Institute of Technology in Stockholm.

\vspace{2mm}\noindent
{\em Software and Data Availability.} The source code used for
the simulations of this study, the {\sc Pencil Code} \citep{PCcollab},
is freely available on \url{https://github.com/pencil-code/}.
The DOI of the code is https://doi.org/10.5281/zenodo.2315093 \citep{v2018.12.16}.
The simulation setups and corresponding data are freely available on
\url{https://doi.org/10.5281/zenodo.4284439} \citep{JB20b}.
\end{acknowledgements}


\end{document}